\documentclass[letterpaper, 10 pt, conference]{ieeeconf}  

\IEEEoverridecommandlockouts                              

\usepackage{graphics} 
\usepackage{amsmath} 
\usepackage{amssymb}  
\usepackage{float}
\usepackage{tcolorbox}
\usepackage{caption}
\usepackage{subcaption}

\newtheorem{prop}{Proposition}
\title{\LARGE \bf
Discrete Optimal Control of Interconnected Mechanical Systems
}

\author{Siddharth H. Nair$^{1}$ and Ravi N. Banavar$^{2}$
\thanks{$^{1}$Siddharth H. Nair is with the Department of Mechanical Engineering at the University of California, Berkeley, USA.    {\tt\small siddharth\_nair@berkeley.edu}}%
\thanks{$^{2}$Ravi N. Banavar with the with the Faculty of Systems and Control Engineering at the Indian Institute of Technology Bombay, Mumbai, India       
{\tt\small banavar@iitb.ac.in }}%
}

\begin{document}

\maketitle
\thispagestyle{empty}
\pagestyle{empty}

\begin{abstract}
This article develops variational integrators for a class of underactuated mechanical systems using the theory of discrete mechanics. Further, a discrete optimal control problem is formulated for the considered class of systems and subsequently solved using variational principles again, to obtain necessary conditions that characterise optimal trajectories. The proposed approach is demonstrated on benchmark underactuated systems and accompanied by numerical simulations.

\end{abstract}

\section{INTRODUCTION}
A particular problem in optimal control of mechanical systems involves transferring a system from its current state to a desired state while minimizing a cost function (like control effort or time). There are two standard methods of approaching this problem. In the first method, the equations of motion are derived using variational principles. Then these differential equations are discretized and applied as algebraic constraints to a nonlinear optimization program to obtain the minimum cost trajectory. The second method involves obtaining optimality conditions for the continuous time system using Pontyragin's maximum principle and then discretize the same to be iteratively solved for an approximate numerical solution (\cite{11}). \\
\par
A more recent method uses the theory of discrete mechanics (\cite{west}) wherein variational principles are used to reformulate Lagrangian and Hamiltonian mechanics in a discrete setting right from the outset, to characterise a set of discrete trajectories. Of these trajectories, the ``optimal" ones are sought by another variational problem that minimizes the cost function. This is called the Discrete Mechanics and Optimal Control (DMOC) method (\cite{sina,12,13,14,tlgoc}) and solutions obtained via this method have been shown to preserve some of the invariants of mechanical systems, such as energy, momentum or the symplectic form.\\
\par
In this article, we seek to find a sequence of control inputs for point-to-point state transfer for a class of underactuated mechanical systems, namely, interconnected mechanical systems (\cite{dhsmfbr, imc1, imc2, nair}). We employ the techniques of discrete mechanics to cast the problem as solutions to a two-point boundary value problem and obtain conditions that necessarily characterise them. Our main contributions are developing variational integrators for interconnected mechanical systems by exploiting their geometric structure and obtaining conditions for computing optimal trajectories for these mechanical systems for any given Lagrangian.\\
\par
The remainder of the article is organised as follows. Section II goes over the basic ingredients that help set up the variational problem and details the process of deriving Lie Group integrators, which leads us to the integrators of interconnected mechanical systems in section III. Section IV formulates  the discrete optimal control problem and provides a set of necessary conditions that characterise the solutions. Sections V and VI demonstrate the proposed approach for the ball and beam system, and the inverted pendulum on a cart respectively, via numerical simulations. Section VII presents concluding remarks and directions for future work.

\section{VARIATIONAL INTEGRATORS FOR LIE GROUPS}
Consider a mechanical system evolving on a matrix Lie group $G$, with its state trajectories evolving on the tangent bundle $TG$ trivialised to $G\times\mathfrak{g}$ using the group action, $\mathfrak{g}$ being the lie algebra of $G$. Defining the Lagrangian of the system as $\mathcal{L}(g,\xi): G\times\mathfrak{g}\rightarrow \mathbb{R}$ and a generalized force $u :\mathbb{R}\rightarrow \mathfrak{g}^*$, the trajectories of the mechanical system are given by the forced Euler-Lagrange equations (\cite{blochbook})
\small
$$\dfrac{d}{dt}D_\xi \mathcal{L}(g,\xi)-\textrm{ad}^*_\xi\cdot D_\xi \mathcal{L}(g,\xi)-T^*_eL_g\cdot D_g\mathcal{L}(g,\xi)=u$$
$$\dot{g}=T_eL_g\cdot\xi=g\xi$$
\normalsize
where $T_eL_g\cdot$ is the lifted left group action.
In the discrete setting, the state trajectories evolve on $G\times G$. A configuration is updated using the group action so as to ensure that the subsequent configurations remain on the Lie group. Define $f_k\in G$ such that 
$$g_{k+1}=R_{f_k}\cdot g_k=g_kf_k$$ where $R_{f_k}\cdot$ is the right group action and the sequence $\{g_k\}^N_{k=0}$ is the discrete flow of the system on $G$ with $g_k$ being the configuration at $t=t_0+kh$ for a fixed time step $h$. Given a discrete Lagrangian $\mathcal{L}_d(g_k,f_k) : G\times G\rightarrow \mathbb{R}$, the action sum is defined as 
\small
$$\mathfrak{A}_d=\sum_{k=0}^{N-1}\mathcal{L}_d(g_k,f_k)$$
\normalsize
For an unforced system, the discrete Hamilton's principle yields the sequence $\{g_k\}^N_{k=0}$ as follows
\small
\begin{align}\label{action}
&\delta \mathfrak{A}_d=\sum^{N-1}_{k=0}\delta \mathcal{L}_d(g_k,f_k)=0\nonumber\\
&\Rightarrow\sum^{N-1}_{k=0}D_{g_k}\mathcal{L}_d(g_k,f_k)\cdot \delta g_k+D_{f_k}\mathcal{L}_d(g_k,f_k)\cdot \delta f_k=0
\end{align}
\normalsize
To compress notation, we denote $\mathcal{L}_d(g_k, f_k) \equiv \mathcal{L}_{dk}$ for the remainder of this paper.
The variation $\delta g_k$ is obtained by considering a one-parameter subgroup on $G$ given by
$g^\epsilon_k=g_k \textrm{exp}(\epsilon\eta_k)$ where $\eta_k\in\mathfrak{g}$ and $\eta_0=\eta_N=0$ for keeping the end points fixed.
\small
\begin{align}\label{vg}
\delta g_k=\left.\dfrac{d}{d\epsilon}\right|_{\epsilon=0}g^\epsilon_k=T_eL_{g_k}\eta_k=g_k\eta_k
\end{align}
\normalsize
The variation of $f_k$ is obtained as follows
\small
\begin{align}\label{vf}
\delta f_k&=\left. \dfrac{d}{d\epsilon}\right|_{\epsilon=0}(g^\epsilon_k)^{-1}g^\epsilon_{k+1}\nonumber\\
&=\left. \dfrac{d}{d\epsilon}\right|_{\epsilon=0}\textrm{exp}(-\epsilon\eta_k)g^{-1}_kg_{k+1}\textrm{exp}(\epsilon\eta_{k+1})\nonumber\\
&=-T_eR_{f_k}\eta_k+T_eL_{f_k}\eta_{k+1}\nonumber\\
&=T_eL_{f_k}\cdot\left\{-Ad_{f^{-1}_k}\eta_k+\eta_{k+1}\right\}
\end{align}
\normalsize
Substituting (2) and (3) into (1) gives us the following equation after taking the adjoints of the 
operators $T_eL_{g_k}$ and $T_eL_{f_k}$
\footnotesize
\begin{align}\label{vareq}
&\sum_{k=0}^{N-1}\langle T^*_eL_{f_k}\cdot D_{f_k}\mathcal{L}_{dk},-Ad_{f^{-1}_k}\eta_k+\eta_{k+1}\rangle\nonumber\\&+\langle T^*_eL_{g_k}\cdot D_{g_k}\mathcal{L}_{dk},\eta_k\rangle=0\nonumber\\
\Rightarrow&\sum_{k=0}^{N-1}\langle T^*_eL_{g_k}\cdot D_{g_k}\mathcal{L}_{dk}-Ad^*_{f^{-1}_k}(T^*_eL_{f_k}\cdot D_{f_k}\mathcal{L}_{dk}),\eta_k\rangle\nonumber\\&+\langle T^*_eL_{f_k}\cdot D_{f_k}\mathcal{L}_{dk},\eta_{k+1}\rangle=0\nonumber\\
\Rightarrow&\sum_{k=1}^{N-1}\langle T^*_eL_{g_k}\cdot D_{g_k}\mathcal{L}_{dk}-Ad^*_{f^{-1}_k}(T^*_eL_{f_k}\cdot D_{f_k}\mathcal{L}_{dk})\nonumber\\&+T^*_eL_{f_{k-1}}\cdot D_{f_{k-1}}\mathcal{L}_{dk-1},\eta_k\rangle=0\hspace{1cm}(\because \eta_0=\eta_N=0)
\end{align}

\normalsize
For all admissible variations, equation (\ref{vareq})  gives us the discrete Euler-Lagrange equations on $G$ as
\small
\begin{subequations}
\begin{align}
T^*_eL_{f_{k-1}}\cdot D_{f_{k-1}}\mathcal{L}_{dk-1}-&Ad^*_{f^{-1}_k}\cdot (T^*_eL_{f_k}\cdot D_{f_k}\mathcal{L}_{dk})\nonumber\\&+T^*_eL_{g_k}\cdot D_{g_k}\mathcal{L}_{dk}=0\\
g_{k}=g_{k-1}f_{k-1}
\end{align}
\end{subequations}
\normalsize
To obtain the forced variant of the discrete Euler-Lagrange equations, we seek to approximate the virtual work done by an external control $U$ when perturbed by a variation, when expressed as follows
\small
$$W=\int_0^TU\cdot\delta g dt=\int_0^T (T^*L_g\cdot U)\cdot \eta dt=\int_0^T u\cdot\eta dt $$
\normalsize
The discrete generalized forces $u^+_{k}, u^-_{k} \in \mathfrak{g}^*$ are chosen such that they approximate the virtual work
$$\int^{t_{k+1}}_{t_k}u\cdot\eta dt\approx u^-_k\cdot\eta_k + u^+_k\cdot\eta_{k+1}$$
Using D' Alembert's principle, the forced discrete Euler-Lagrange equations are then given by 

\small
\begin{subequations}
\begin{align}
&T^*_eL_{f_{k-1}}\cdot D_{f_{k-1}}\mathcal{L}_{dk-1}-Ad^*_{f^{-1}_k}(T^*_eL_{f_k}\cdot D_{f_k}\mathcal{L}_{dk})\nonumber\\&+T^*_eL_{g_k}\cdot D_{g_k}\mathcal{L}_{dk}+u^-_k+u^+_{k-1}=0\\
&g_{k}=g_{k-1}f_{k-1}
\end{align}
\end{subequations}
\normalsize

The discrete Legendre transforms $\mathbb{F}^+\mathcal{L}_d,\ \mathbb{F}^-\mathcal{L}_d\ :\ G\times G\rightarrow G\times \mathfrak{g}^*$ are given by
\small
\begin{subequations}
\begin{align}
\mathbb{F}^+\mathcal{L}_d(g_k,f_k)&=(g_kf_k,\mu_{k+1})\\
\mathbb{F}^-\mathcal{L}_d(g_k,f_k)&=(g_k,\mu_{k})
\end{align}
\end{subequations}
\normalsize
where $\mu_k$ and $\mu_{k+1}$ are given by
\small
\begin{subequations}
\begin{align}
&\mu_k=-T^*_eL_{g_k}\cdot D_{g_k}\mathcal{L}_{d}+\textrm{Ad}^*_{f^{-1}_k}(T^*_eL_{f_k}\cdot D_{f_k}\mathcal{L}_d)-u^-_k\\
&\mu_{k+1}=T^*_eL_{f_k}\cdot D_{f_k}\mathcal{L}_d +u^+_{k}
\end{align}
\end{subequations}
\normalsize
The discrete Legendre transforms thus give us the discrete-time Hamilton's equations as
\small
\begin{subequations}
\begin{align}
&\mu_k=-T^*_eL_{g_k}\cdot D_{g_k}\mathcal{L}_{d}+\textrm{Ad}^*_{f^{-1}_k}(T^*_eL_{f_k}\cdot D_{f_k}\mathcal{L}_d)-u^-_k\\
&g_{k+1}=g_kf_k\\
&\mu_{k+1}=\textrm{Ad}^*_{f_k}(\mu_k+T^*_eL_{g_k}\cdot D_{g_k}\mathcal{L}_d +u^-_{k})+u^+_{k}
\end{align}
\end{subequations}
\normalsize
\section{INTERCONNECTED MECHANICAL SYSTEMS}

In this work, we consider interconnected mechanical systems whose configurations variables can be expressed as $G \ni g_k=(g_{ak},g_{uk}) \in G_a\times G_u$ where the $g_{ak}$ are the actuated variables belonging to lie group $G_a$ and $g_{uk}$ are the unactuated variables belonging to lie group $G_u$. $G \ni f_k=(f_{ak},f _{uk}) \in G_a\times G_u$ is decomposed similarly. Using the product structure of the configuration space $G$, the virtual work can be approximated solely in terms of the generalized inputs and lie algebraic elements in $\mathfrak{g}_a$ to give the following discrete Euler-Lagrange equations
\small
\begin{subequations}
\label{clua}
\begin{align}
&T^*_eL_{f_{ak-1}}\cdot D_{f_{ak-1}}\mathcal{L}_{dk-1}-Ad^*_{f^{-1}_{ak}}(T^*_eL_{f_{ak}}\cdot D_{f_{ak}}\mathcal{L}_{dk})\nonumber\\&+T^*_eL_{g_{ak}}\cdot D_{g_{ak}}\mathcal{L}_{dk}+u^-_k+u^+_{k-1}=0\\
&T^*_eL_{f_{uk-1}}\cdot D_{f_{uk-1}}\mathcal{L}_{dk-1}-Ad^*_{f^{-1}_{uk}}(T^*_eL_{f_{uk}}\cdot D_{f_{uk}}\mathcal{L}_{dk})\nonumber\\&+T^*_eL_{g_{uk}}\cdot D_{g_{uk}}\mathcal{L}_{dk}=0\\
&g_{ak}=g_{ak-1}f_{ak-1}\\
&g_{uk}=g_{uk-1}f_{uk-1}
\end{align}
\end{subequations}
\normalsize
The discrete Hamilton's equations are thus given by 
\small
\begin{subequations}
\label{cluah}
\begin{align}
&\mu_{ak}=-T^*_eL_{g_{ak}}\cdot D_{g_{ak}}\mathcal{L}_{dk}+\textrm{Ad}^*_{f^{-1}_{ak}}(T^*_eL_{f_{ak}}\cdot D_{f_{ak}}\mathcal{L}_{dk})-u^-_k\\
&g_{ak+1}=g_{ak}f_{ak}\\
&\mu_{ak+1}=\textrm{Ad}^*_{f_{ak}}(\mu_{ak}+T^*_eL_{g_{ak}}\cdot D_{g_{ak}}\mathcal{L}_d +u^-_{k})+u^+_{k}\\
&\mu_{uk}=-T^*_eL_{g_{uk}}\cdot D_{g_{uk}}\mathcal{L}_{dk}+\textrm{Ad}^*_{f^{-1}_{uk}}(T^*_eL_{f_{uk}}\cdot D_{f_{uk}}\mathcal{L}_{dk})\\
&g_{uk+1}=g_{uk}f_{uk}\\
&\mu_{uk+1}=\textrm{Ad}^*_{f_{uk}}(\mu_{uk}+T^*_eL_{g_{uk}}\cdot D_{g_{uk}}\mathcal{L}_{dk})
\end{align}
\end{subequations}
\normalsize

\section{DISCRETE OPTIMAL CONTROL PROBLEM}
We consider the problem of deriving a sequence optimal control inuputs for point-to-point transfer of systems states with discrete time dynamics described by (\ref{cluah}) over a fixed horizon $N$. In the sequel, we use the trapezoidal rule to approximate the control as follows
\begin{subequations}
\begin{align}
u^-_k&=\frac{1}{2}hu(t_0+kh)=\frac{1}{2}hu_k\\
u^+_k&=\frac{1}{2}hu(t_0+(k+1)h)=\frac{1}{2}hu_{k+1}
\end{align}
\end{subequations}
 Let the cost functional $\mathcal{J}_d$ be of the form
\begin{align}
\mathcal{J}_d=\sum_{k=0}^{N-1}\phi_d(g_k,f_k,u_k)
\end{align}
where $\phi_d : G\times G \times \mathfrak{g}^*\rightarrow \mathbb{R}$ is the cost-per-stage for each $k=0,1,.,N-1$. Given initial conditions $(g_0,\mu_0)$ and terminal conditions $(g^f, \mu^f)$, the discrete-time optimal control problem is given by

\begin{align}\label{docp}
&\textrm{Given } N, g_0, \mu_0\nonumber\\
&\textrm{min}_{u_k}(\mathcal{J}_d=\sum_{k=0}^{N-1}\phi_d(g_k,f_k,u_k))\nonumber\\
&\textrm{such that } g_N=g^f,\ \mu_N=\mu^f, \nonumber\\
&\textrm{subject to } (\ref{cluah})
\end{align}

The cost functional can be augmented using Lagrange multipliers $\lambda^1_k, \lambda^3_k\in\mathfrak{g}_a$, $\lambda^4_k, \lambda^6_k \in \mathfrak{g}_u$, $\lambda^2_k\in\mathfrak{g}^*_a$ and $\lambda^5_k \in \mathfrak{g}^*_u$ as follows
\small
\begin{align}
&\mathcal{J}_d=\sum_{k=0}^{N-1}\mathcal{J}_{d0k}+\mathcal{J}_{d1k}+\mathcal{J}_{d2k}+\mathcal{J}_{d3k}+\mathcal{J}_{d4k}+\mathcal{J}_{d5k}+\mathcal{J}_{d6k}
\end{align}
\normalsize
where
\small
\begin{align*}
\mathcal{J}_{d0k}=&\phi_d(g_k,f_k,u_k)\\
\mathcal{J}_{d1k}=&\langle T^*_eL_{g_{ak}}\cdot D_{g_{ak}}\mathcal{L}_{dk}-\textrm{Ad}^*_{f^{-1}_{ak}}(T^*_eL_{f_{ak}}\cdot D_{f_{ak}}\mathcal{L}_{dk})\\&+\frac{1}{2}hu_k+\mu_{ak}, \lambda^1_k\rangle\\
\mathcal{J}_{d2k}=&\langle\lambda^2_k, \textrm{log}(g^{-1}_{ak}g_{ak+1})-\textrm{log}(f_{ak})\rangle\\
\mathcal{J}_{d3k}=&\langle -\textrm{Ad}^*_{f_{ak}}(\mu_{ak}+T^*_eL_{g_{ak}}\cdot D_{g_{ak}}\mathcal{L}_{dk}+\frac{1}{2}hu_k)\\&-\frac{1}{2}hu_{k+1}+\mu_{ak+1}, \lambda^3_k\rangle\\
\mathcal{J}_{d4k}=&\langle T^*_eL_{g_{uk}}\cdot D_{g_{uk}}\mathcal{L}_{dk}-\textrm{Ad}^*_{f^{-1}_{uk}}(T^*_eL_{f_{uk}}\cdot D_{f_{uk}}\mathcal{L}_{dk})\\&+\mu_{uk}, \lambda^4_k\rangle\\
\mathcal{J}_{d5k}=&\langle\lambda^5_k, \textrm{log}(g^{-1}_{uk}g_{uk+1})-\textrm{log}(f_{uk})\rangle\\
\mathcal{J}_{d6k}=&\langle \mu_{uk+1}-(\textrm{Ad}^*_{f_{uk}}\cdot(\mu_{uk}+T^*_eL_{g_{uk}}\cdot D_{g_{uk}}\mathcal{L}_{dk})), \lambda^6_k\rangle
\end{align*}
\normalsize
\\
\textbf{Key Assumption:} The $\textrm{log} : G\rightarrow \mathfrak{g}$ map is well-defined when $f_{ak},\  g^{-1}_{ak}g_{ak+1},\ f_{uk},\ g^{-1}_{uk}g_{uk+1}$ are close to the identity element on $G$. We assume that a sufficiently small time step $h$ is chosen to accomplish this.\\\\
We obtain the necessary conditions for optimality using calculus of variations. Discrete-time Hamiltion's principle gives us
\begin{equation}
\label{vj1}
\delta \mathcal{J}_{d}=\delta \mathcal{J}_{d0}+\delta \mathcal{J}_{d1}+\delta \mathcal{J}_{d2}+\delta \mathcal{J}_{d3}+\delta \mathcal{J}_{d4}+\delta \mathcal{J}_{d5}+\delta \mathcal{J}_{d6}=0
\end{equation}
To obtain the derivatives of the $log$ maps in $\mathcal{J}_{d2}$ and $\mathcal{J}_{d5}$, we use the BCH formula since we are considering matrix lie groups.\\\\
\textit{Baker-Campbell-Hausdorff formula}\\\\
Let $X$ and $Y$ be elements of a Lie algebra $\mathfrak{g}$ of some Matrix Lie group $G$ with the exponential map, $\exp: \mathfrak{g}\rightarrow G$. Then for $\exp(X)$, $\exp(Y)$ close to the identity element of $G$, the $\exp$ map is a diffeomorphism with its inverse $\log : G\rightarrow\mathfrak{g}$ given by the following 
\small
\begin{align}
\label{bch1}
\log(\exp(X)\exp(Y))=X+\dfrac{\textrm{ad}_X\exp(\textrm{ad}_X)}{\exp(\textrm{ad}_X)-1}Y+O(Y^2)
\end{align}
\normalsize
and 
\small
\begin{align}
\label{bch2}
\log(\exp(X)\exp(Y))&=X+Y+\frac{1}{2}[X,Y]\nonumber\\&+\frac{1}{12}([X,[X,Y]]+[Y,[Y,X]])+\dots
\end{align}
\normalsize
We now state our main result and defer the proof to the appendix.
\subsection*{\textbf{Necessary Conditions for Optimality}}
\textit{Given an interconnected mechanical system governed by discrete dynamics given by (\ref{cluah}), the trajectories of the system from initial condition $(g_0,\mu_0)$ to terminal condition $(g^f, \mu^f)$ that minimize the cost functional $\mathcal{J}_d=\sum_{k=0}^{N-1}\phi_d(g_k,f_k,u_k))$ necessarily satisfy the following equations.}
%
%
\footnotesize
\begin{subequations}
\label{doce1}
\begin{flalign}
&\left\{\textbf{Multiplier (Adjoint) Equations}\right\}\nonumber\\
&T^*_eL_{g_{ak}}\cdot D_{g_{ak}}\phi_d+\mathcal{M}^{ag}_{ag}(\lambda^1_k-\textrm{Ad}_{f_{ak}}\lambda^3_k)-\mathcal{M}^{ag}_{af}(\textrm{Ad}_{f^{-1}_{ak}}\lambda^1_k)\nonumber\\&
+\mathcal{M}^{ag}_{ug}(\lambda^4_k-\textrm{Ad}_{f_{uk}}\lambda^6_k)-\mathcal{M}^{ag}_{uf}(\textrm{Ad}_{f^{-1}_{uk}}\lambda^4_k)=-\lambda^2_{k-1}+\textrm{Ad}^*_{f^{-1}_{ak}}\lambda^2_k\\
\nonumber\\
&T^*_eL_{g_{uk}}\cdot D_{g_{uk}}\phi_d+\mathcal{M}^{ug}_{ag}(\lambda^1_k-\textrm{Ad}_{f_{ak}}\lambda^3_k)-\mathcal{M}^{ug}_{af}(\textrm{Ad}_{f^{-1}_{ak}}\lambda^1_k)\nonumber\\&
+\mathcal{M}^{ug}_{ug}(\lambda^4_k-\textrm{Ad}_{f_{uk}}\lambda^6_k)-\mathcal{M}^{ug}_{uf}(\textrm{Ad}_{f^{-1}_{uk}}\lambda^4_k)=-\lambda^5_{k-1}+\textrm{Ad}^*_{f^{-1}_{uk}}\lambda^5_k\\
\nonumber\\
&T^*_eL_{f_{ak}}\cdot D_{f_{ak}}\phi_d+\mathcal{M}^{af}_{ag}(\lambda^1_k-\textrm{Ad}_{f_{ak}}\lambda^3_k)-\mathcal{M}^{af}_{af}(\textrm{Ad}_{f^{-1}_{ak}}\lambda^1_k)\nonumber\\& -\textrm{ad}^*_{\textrm{Ad}_{f_{ak}^{-1}}\lambda^1_k}(M_{af})+\textrm{Ad}^*_{f_{ak}}\textrm{ad}^*_{\textrm{Ad}_{f_{ak}}\lambda^3_k}(\mu_{ak}+M_{ag}+\frac{1}{2}hu_k)\nonumber\\&
+\mathcal{M}^{af}_{ug}(\lambda^4_k-\textrm{Ad}_{f_{uk}}\lambda^6_k)-\mathcal{M}^{af}_{uf}(\textrm{Ad}_{f^{-1}_{uk}}\lambda^4_k)=\lambda^2_{k}\\
\nonumber\\
&T^*_eL_{f_{uk}}\cdot D_{f_{uk}}\phi_d+\mathcal{M}^{uf}_{ag}(\lambda^1_k-\textrm{Ad}_{f_{ak}}\lambda^3_k)-\mathcal{M}^{uf}_{af}(\textrm{Ad}_{f^{-1}_{ak}}\lambda^1_k)\nonumber\\&-\textrm{ad}^*_{\textrm{Ad}_{f_{uk}^{-1}}\lambda^4_k}(M_{uf})+\textrm{Ad}^*_{f_{uk}}\textrm{ad}^*_{\textrm{Ad}_{f_{uk}}\lambda^6_k}(\mu_{uk}+M_{ug})\nonumber\\&
+\mathcal{M}^{uf}_{ug}(\lambda^4_k-\textrm{Ad}_{f_{uk}}\lambda^6_k)-\mathcal{M}^{uf}_{uf}(\textrm{Ad}_{f^{-1}_{uk}}\lambda^4_k)=\lambda^5_{k}\\
\nonumber\\
&\lambda^1_k-\textrm{Ad}^*_{f_{ak}}\lambda^3_k=-\lambda^3_{k-1}\\
\nonumber\\
&\lambda^4_k-\textrm{Ad}^*_{f_{uk}}\lambda^6_k=-\lambda^6_{k-1}\hspace{1cm}\forall k=1,2,..N-1
\end{flalign}
\end{subequations}
\begin{subequations}
\label{doce2}
\begin{flalign}
&\left\{\textbf{Optimality Equations}\right\}\hspace{12cm}\nonumber\\
&D_{u_{k}}\phi_d+\frac{h}{2}\lambda^1_k-\frac{h}{2}(\lambda^3_{k-1}+\textrm{Ad}_{f_{ak}}\lambda^3_k)=0\hspace{1cm}\forall k=1,2,..N-1\\
&D_{u_{0}}\phi_d+\frac{h}{2}\lambda^1_0-\frac{h}{2}\textrm{Ad}_{f_{a0}}\lambda^3_0=0
\end{flalign}
\end{subequations}
\begin{subequations}
\label{doce3}
\begin{flalign}
&\left\{\textbf{Boundary Conditions}\right\}\hspace{12cm}\nonumber\\
&g_N=g^f\quad \mu_N=\mu^f
\end{flalign}
\end{subequations}
\begin{subequations}
\label{doce4}
\begin{flalign}
&\left\{\textbf{State Equations}\right\}\hspace{12cm}\nonumber\\
&\mu_{ak}=-T^*_eL_{g_{ak}}\cdot D_{g_{ak}}\mathcal{L}_{dk}+\textrm{Ad}^*_{f^{-1}_{ak}}(T^*_eL_{f_{ak}}\cdot D_{f_{ak}}\mathcal{L}_{dk})-u^-_k\\
\nonumber\\
&g_{ak+1}=g_{ak}f_{ak}\\
\nonumber\\
&\mu_{ak+1}=\textrm{Ad}^*_{f_{ak}}(\mu_{ak}+T^*_eL_{g_{ak}}\cdot D_{g_{ak}}\mathcal{L}_{dk} +u^-_{k})+u^+_{k}\\
\nonumber\\
&\mu_{uk}=-T^*_eL_{g_{uk}}\cdot D_{g_{uk}}\mathcal{L}_{dk}+\textrm{Ad}^*_{f^{-1}_{uk}}(T^*_eL_{f_{uk}}\cdot D_{f_{uk}}\mathcal{L}_{dk})\\
\nonumber\\
&g_{uk+1}=g_{uk}f_{uk}\\
\nonumber\\
&\mu_{uk+1}=\textrm{Ad}^*_{f_{uk}}(\mu_{uk}+T^*_eL_{g_{uk}}\cdot D_{g_{uk}}\mathcal{L}_{dk})\hspace{1cm}\forall k=0,1,..N-1
\end{flalign}
\end{subequations}
\normalsize
%
%
\textit{where the derivatives of the Lagrangian with respect to various group elements are defined as follows}
\small
\begin{subequations}
\label{M}
\begin{align}
M_{ag}(g_k,f_k)&=T^*_eL_{g_{ak}}\cdot D_{g_{ak}}\mathcal{L}_{dk}\\
M_{af}(g_k,f_k)&=T^*_eL_{f_{ak}}\cdot D_{f_{ak}}\mathcal{L}_{dk}\\
M_{ug}(g_k,f_k)&=T^*_eL_{g_{uk}}\cdot D_{g_{uk}}\mathcal{L}_{dk}\\
M_{uf}(g_k,f_k)&=T^*_eL_{f_{uk}}\cdot D_{f_{uk}}\mathcal{L}_{dk}
\end{align}
\end{subequations}
\normalsize
\textit{while the second derivatives of the Lagrangian with respect to the various group elements, acting linearly on the Lagrange multipliers are defined as follows}
\footnotesize
\begin{subequations}
\label{calM}
\begin{align}
&\mathcal{M}^{ag}_{ag} : G\times G\times \mathfrak{g}_{a}\rightarrow \mathfrak{g}^*_a\nonumber\\
&\langle D_{g_{ak}}M_{ag}\cdot\delta g_{ak},\lambda_k\rangle =\langle D_{g_{ak}}M_{ag}\cdot T_eL_{g_{ak}}\eta_{ak},\lambda_k\rangle=\langle \mathcal{M}^{ag}_{ag}(\lambda_k),\eta_{ak}\rangle\\
&\mathcal{M}^{ug}_{ag} : G\times G\times \mathfrak{g}_{a}\rightarrow \mathfrak{g}^*_u\nonumber\\
&\langle D_{g_{uk}}M_{ag}\cdot\delta g_{uk},\lambda_k\rangle =\langle D_{g_{uk}}M_{ag}\cdot T_eL_{g_{uk}}\eta_{uk},\lambda_k\rangle=\langle \mathcal{M}^{ug}_{ag}(\lambda_k),\eta_{uk}\rangle\\
&\mathcal{M}^{af}_{ag} : G\times G\times \mathfrak{g}_{a}\rightarrow \mathfrak{g}^*_a\nonumber\\
&\langle D_{f_{ak}}M_{ag}\cdot\delta f_{ak},\lambda_k\rangle =\langle D_{f_{ak}}M_{ag}\cdot T_eL_{f_{ak}}\chi_{ak},\lambda_k\rangle=\langle \mathcal{M}^{af}_{ag}(\lambda_k),\chi_{ak}\rangle\\
&\mathcal{M}^{uf}_{ag} : G\times G\times \mathfrak{g}_{a}\rightarrow \mathfrak{g}^*_u\nonumber\\
&\langle D_{f_{uk}}M_{ag}\cdot\delta f_{uk},\lambda_k\rangle =\langle D_{f_{uk}}M_{ag}\cdot T_eL_{f_{uk}}\chi_{uk},\lambda_k\rangle=\langle \mathcal{M}^{uf}_{ag}(\lambda_k),\chi_{uk}\rangle
\end{align}
\end{subequations}
\normalsize
\textit{We similarly define functions} $\mathcal{M}^{ag}_{af}, \mathcal{M}^{ug}_{af}, \mathcal{M}^{af}_{af},$ $\mathcal{M}^{uf}_{af}, \mathcal{M}^{ag}_{ug}, \mathcal{M}^{ug}_{ug}, \mathcal{M}^{af}_{ug},\mathcal{M}^{uf}_{ug}, \mathcal{M}^{ag}_{uf}, \mathcal{M}^{ug}_{uf}, \mathcal{M}^{af}_{uf}$ and $\mathcal{M}^{uf}_{uf}$\\\\ 
In the next section we compute the necessary conditions for a ball and beam system to demonstrate the proposed control synthesis strategy.


\section{EXAMPLE: Ball and Beam System}
Consider a ball of mass $m_b$ sliding along a beam with mass $m_r$. The situation is described in the figure below.
\begin{figure}[H]
\centering
\includegraphics[scale=0.4]{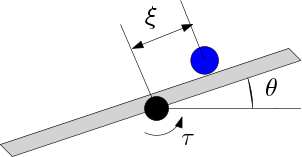}
\caption{Ball and beam system}
\end{figure}
$\theta$ is the angle that the beam makes with the horizontal while $\xi$ is the position of the ball on the beam, measured from the beam's pivot. The beam is actuated by a motor at it's pivot. The system trajectories evolve over the manifold $Q=S^1\times \mathbb{R}$. The beam constitutes the actuated subsystem while the ball constitutes the unactuated subsystem. The Lagrangian of the system is given by
\small
\begin{align*}
&\mathcal{L}\ :\ TQ\rightarrow\mathbb{R}\\
&\mathcal{L} (\xi,\theta,\dot{\xi},\dot{\theta})= \frac{1}{2}I_r\dot{\theta}^2+\frac{1}{2}m_b(\dot{\xi}^2+\xi^2\dot{\theta}^2)-m_bg\xi\sin\theta
\end{align*}
\normalsize
We now proceed to discretize the system dynamics using the trapezoidal rule. For a time step $h$, the discrete Lagrangian is given by
\small
\begin{align*}
\mathcal{L}_{dk}\ &:\ Q\times Q\rightarrow \mathbb{R}\\
\mathcal{L}_{dk}(\xi_k,\theta_k,\Delta \xi_k, \Delta \theta_k)&=hL(\xi_k,\theta_k, \frac{\Delta \xi_k}{h}, \frac{\Delta \theta_k}{h})\\
&=\frac{1}{2h}I_r\Delta \theta_k^2+\frac{1}{2h}m_b(\Delta\xi_k^2+\xi_k^2\Delta \theta_k^2)\\&-m_bhg\xi_k\sin\theta_k
\end{align*}
\normalsize
The discrete-time Euler-Lagrange equations of motion are given by
\small
\begin{subequations}
\begin{align}
&\frac{I_r}{h}(\Delta\theta_{k-1}-\Delta\theta_k)+\frac{m_b}{h}(\xi^2_{k-1}\Delta\theta_{k-1}-\xi^2_k\Delta\theta_k)\nonumber\\&-m_bhg\xi_k\cos\theta_k+\frac{h}{2}(u_k+u_{k-1})=0\\
&\frac{m_b}{h}(\Delta\xi_{k-1}-\Delta\xi_k)+\frac{m_b}{h}(\xi_k\Delta\theta_k^2)-m_bhg\sin\theta_k=0\\
&\theta_{k}=\theta_{k-1}+\Delta\theta_{k-1}\\
&\xi_{k}=\xi_{k-1}+\Delta \xi_{k-1}
\end{align}
\end{subequations}
\normalsize

The discrete-time Hamilton's equations are given by
\small
\begin{subequations}
\label{heqbb}
\begin{align}
&\mu_{ak}=m_bhg\xi_k\cos\theta_k+\frac{I_r}{h}\Delta\theta_k+\frac{m_b}{h}\xi^2_k\Delta\theta_k-\frac{h}{2}u_k\\
&\theta_{k+1}=\theta_{k}+\Delta\theta_{k}\\
&\mu_{ak+1}=\mu_{ak}-m_bhg\xi_k\cos\theta_k+\frac{h}{2}u_k+\frac{h}{2}u_{k+1}\\
&\mu_{uk}=\frac{m_b}{h}\Delta\xi_k-\frac{m_b}{h}(\xi_k\Delta\theta_k^2)+m_bhg\sin\theta_k\\
&\xi_{k+1}=\xi_{k}+\Delta \xi_{k}\\
&\mu_{uk+1}=\mu_{uk}+\frac{m_b}{h}(\xi_k\Delta\theta_k^2)-m_bhg\sin\theta_k
\end{align}
\end{subequations}
\normalsize

Let us find the sequence of controls $\{u_k\}$ which minimizes the cost $\mathcal{J}_d=\sum_{k=0}^{N-1}\frac{1}{2}u^2_k$ and gets the system from $(\theta_0,\xi_0,\mu_{a0},\mu_{u0})=(\theta_0,\xi_0,0,0)$ to $(\theta_N,\xi_N,\mu_{aN},\mu_{uN})=(\theta_f,\xi_f,0,0)$.\\
We thus wish to solve the problem 
\begin{align}\label{edocp}
\textrm{min}_{u_k}(\mathcal{J}_d&=\sum_{k=0}^{N-1}\frac{1}{2}u^2_k)\nonumber\\
\textrm{such that }(\theta_0,\xi_0,\mu_{a0},\mu_{u0})&=(\theta_0,\xi_0,0,0)\nonumber\\  (\theta_N,\xi_N,\mu_{aN},\mu_{uN})&=(\theta_f,\xi_f,0,0)\nonumber\\
\textrm{subject to } (\ref{heqbb})
\end{align}
The following table presents the first and second derivatives of the Lagrangian (defined as in (\ref{M}) and (\ref{calM})) in order to obtain the multiplier equations and optimality condition as in (\ref{doce1}) and (\ref{doce2})
\begin{center}
\begin{tabular}{ |c|c|c|c| } 
\hline
 $M_{ag}=-m_bgh\xi_k\cos\theta_k$ & $M_{af}=\frac{I_r}{h}\Delta\theta_k+\frac{m_b}{h}\xi^2_k\Delta\theta_k$\\
 \hline
 $\mathcal{M}^{ag}_{ag}=m_bgh\xi_k\sin\theta_k\lambda$ & $\mathcal{M}^{ag}_{af}=0$\\
$\mathcal{M}^{af}_{ag}=0$ & $\mathcal{M}^{af}_{af}=(\frac{I_r}{h}+\frac{m_b}{h}\xi^2_k)\lambda$\\
$\mathcal{M}^{ug}_{ag}=-m_bgh\cos\theta_k\lambda$ & $\mathcal{M}^{ug}_{af}=2\frac{m_b}{h}\xi_k\Delta\theta_k\lambda$\\
$\mathcal{M}^{uf}_{ag}=0$ & $\mathcal{M}^{uf}_{af}=0$\\
\hline
$M_{ug}=\frac{m_b}{h}\xi_k\Delta\theta_k^2-m_bgh\sin\theta_k$ & $M_{uf}=\frac{m_b}{h}\Delta\xi_k$ \\
\hline
$\mathcal{M}^{ag}_{ug}=-m_bgh\cos\theta_k\lambda$ & $\mathcal{M}^{ag}_{uf}=0$\\
$\mathcal{M}^{af}_{ug}=2\frac{m_b}{h}\xi_k\Delta\theta_k\lambda$ & $\mathcal{M}^{af}_{uf}=0$\\
$\mathcal{M}^{ug}_{ug}=\frac{m_b}{h}\Delta\theta_k^2\lambda$ & $\mathcal{M}^{ug}_{uf}=0$\\
$\mathcal{M}^{uf}_{ug}=0$ & $\mathcal{M}^{uf}_{uf}=\frac{m_b}{h}\lambda$\\\hline
\end{tabular}
\end{center}
\normalsize
%
The multiplier equations and condition of optimality are therefore given by
\small
\begin{subequations}
\begin{align}
&m_bgh\xi_k\sin\theta_k(\lambda^1_k-\lambda^3_k)-m_bgh\cos\theta_k(\lambda^4_k-\lambda^6_k)=-\lambda^2_{k-1}+\lambda^2_k\\
&m_bgh\cos\theta_k(\lambda^3_k-\lambda^1_k)-2\frac{m_b}{h}\xi_k\Delta\theta_k\lambda^1_k+\frac{m_b}{h}\Delta\theta_k^2(\lambda^4_k-\lambda^6_k)\nonumber\\&\hspace{5cm}=-\lambda^5_{k-1}+\lambda^5_k\\
&-(\frac{I_r}{h}+\frac{m_b}{h}\xi^2_k)\lambda^1_k+2\frac{m_b}{h}\xi_k\Delta\theta_k(\lambda^4_k-\lambda^6_k)=\lambda^2_k\\
&-\frac{m_b}{h}\lambda^4_k=\lambda^5_k\\
&\lambda^1_k-\lambda^3_k=-\lambda^3_{k-1}\\
&\lambda^4_k-\lambda^6_k=-\lambda^6_{k-1}\\
&u_k+\frac{h}{2}\lambda^1_k-\frac{h}{2}\lambda^3_k=\frac{h}{2}\lambda^3_{k-1}\hspace{1cm} \forall k=1,2,\dots, N-1\\
&u_0+\frac{h}{2}\lambda^1_0-\frac{h}{2}\lambda^3_0=0
\end{align}
\end{subequations}
\normalsize
The above equations are solved using the multiple shooting method described in \cite{karmvir}.
\subsection*{Simulation Results}
The solution was obtained numerically on a PC by implementing the above algorithm with the following parameters
\begin{center}
 \begin{tabular}{||c c||} 
 \hline
 Parameter & Value  \\ [0.5ex] 
 \hline\hline
 $m_b$ & $0.5\ kg$ \\ 
 \hline
 $I_r$ & $6\ kg\ m^2$ \\
 \hline
 $g$ & $9.8\ m\ s^{-2}$ \\
 \hline
 $h$ & $0.01\ s$ \\
 \hline
 N & 1000\\ [1ex] 
 \hline
\end{tabular}
\end{center}
We present our results for two sets of boundary conditions 
\subsubsection*{\textbf{Case 1}}
The initial and terminal conditions are set as 
\begin{center}
 \begin{tabular}{||c c c c||} 
 \hline\hline
 $\theta_0= $& 0 &  $\theta_N =$& 0 \\ 
 \hline
 $\mu_{a0}=$ & $0\ kg\ m^2\ s^{-1}$ &  $\mu_{aN}=$ & $0\ kg\ m^2\ s^{-1}$ \\
 \hline
 $\xi_0 =$& $0.5\ m$ &  $\xi_N=$ & $0\ m$ \\
 \hline
 $\mu_{u0}=$ & $0\ kg\ m\ s^{-1}$ & $\mu_{uN}=$ & $0\ kg\ m\ s^{-1}$ \\[1ex] 
 \hline\hline
\end{tabular}
\end{center}

\begin{figure}[H]
  \centering
  \includegraphics[scale=0.2]{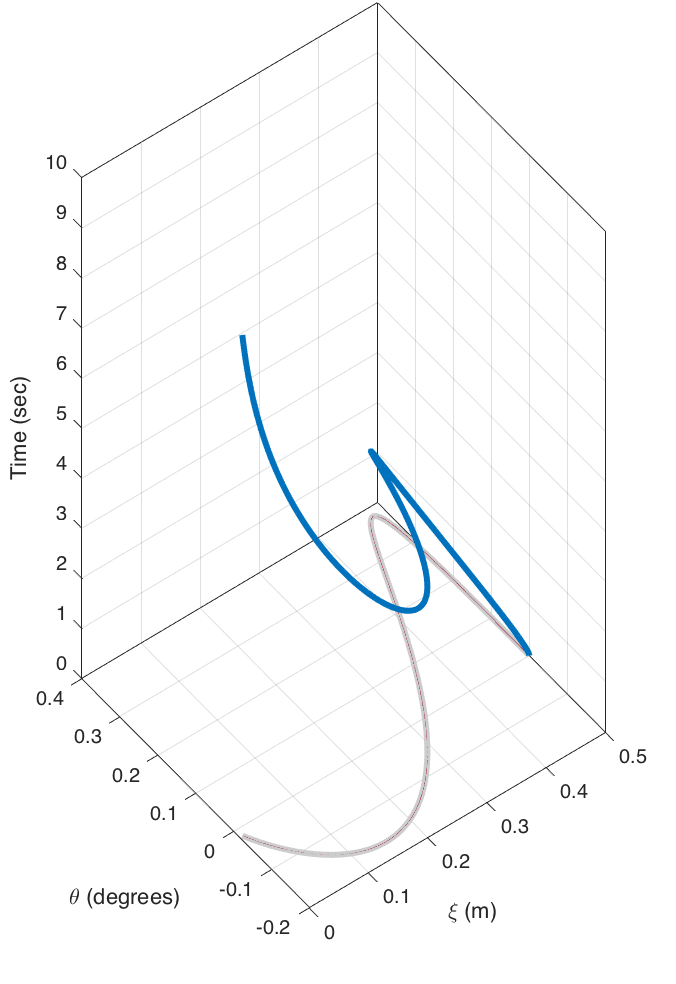}
  \caption{Evolution of ball and beam configurations with time for case 1}
  \label{traj1}
\end{figure}

\begin{figure}[H]
\begin{subfigure}{0.5\textwidth}
  \centering
  \includegraphics[scale=0.18]{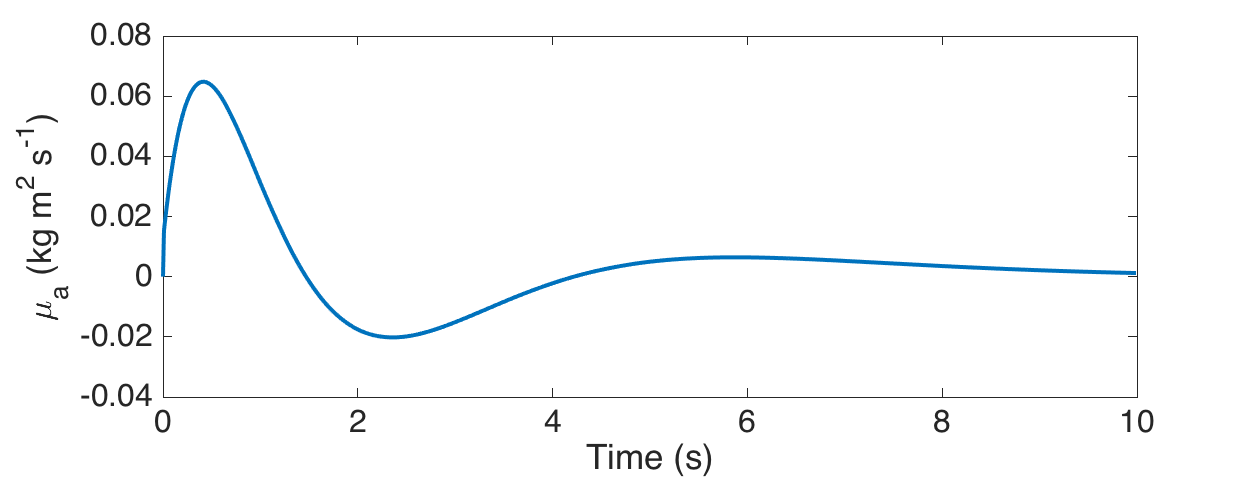}
\end{subfigure}
\begin{subfigure}{0.5\textwidth}
  \centering
  \includegraphics[scale=0.18]{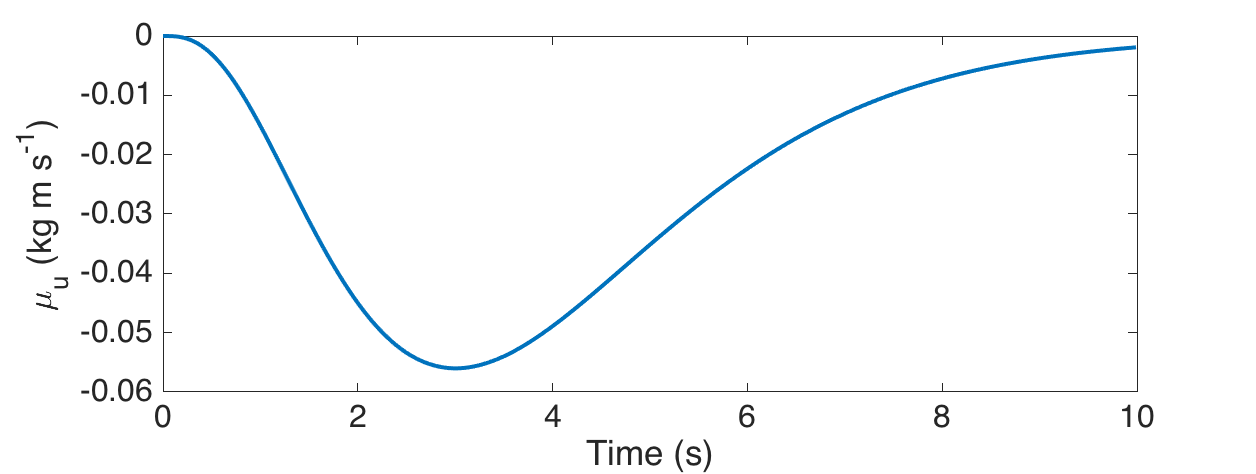}
  \end{subfigure}
  \begin{subfigure}{0.5\textwidth}
  \centering
  \includegraphics[scale=0.18]{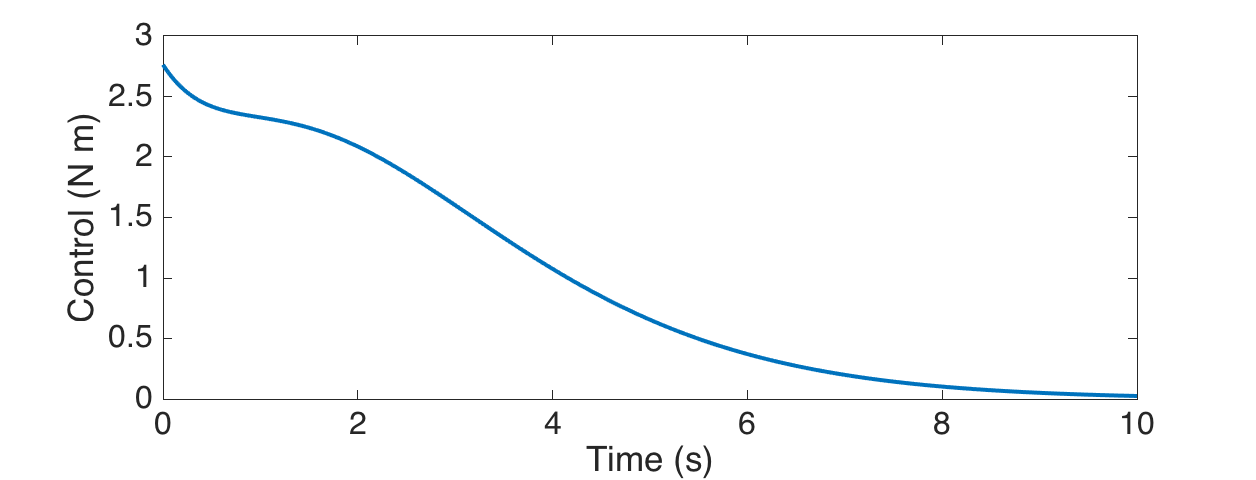}
  \end{subfigure}
  \caption{Momenta and control input versus time for case 1}
\end{figure}
We see that the numerical solution obtained respects the boundary conditions satisfactorily, stabilizing both, the ball and the beam.
\subsubsection*{\textbf{Case 2}}
The initial and terminal conditions are set as 
\begin{center}
 \begin{tabular}{||c c c c||} 
 \hline\hline
 $\theta_0= $& $18^{\circ}$ &  $\theta_N =$& 0 \\ 
 \hline
 $\mu_{a0}=$ & $0\ kg\ m^2\ s^{-1}$ &  $\mu_{aN}=$ & $0\ kg\ m^2\ s^{-1}$ \\
 \hline
 $\xi_0 =$& $0.5\ m$ &  $\xi_N=$ & $0\ m$ \\
 \hline
 $\mu_{u0}=$ & $0\ kg\ m\ s^{-1}$ & $\mu_{uN}=$ & $0\ kg\ m\ s^{-1}$ \\[1ex] 
 \hline\hline
\end{tabular}
\end{center}

\begin{figure}[H]
  \centering
  \includegraphics[scale=0.2]{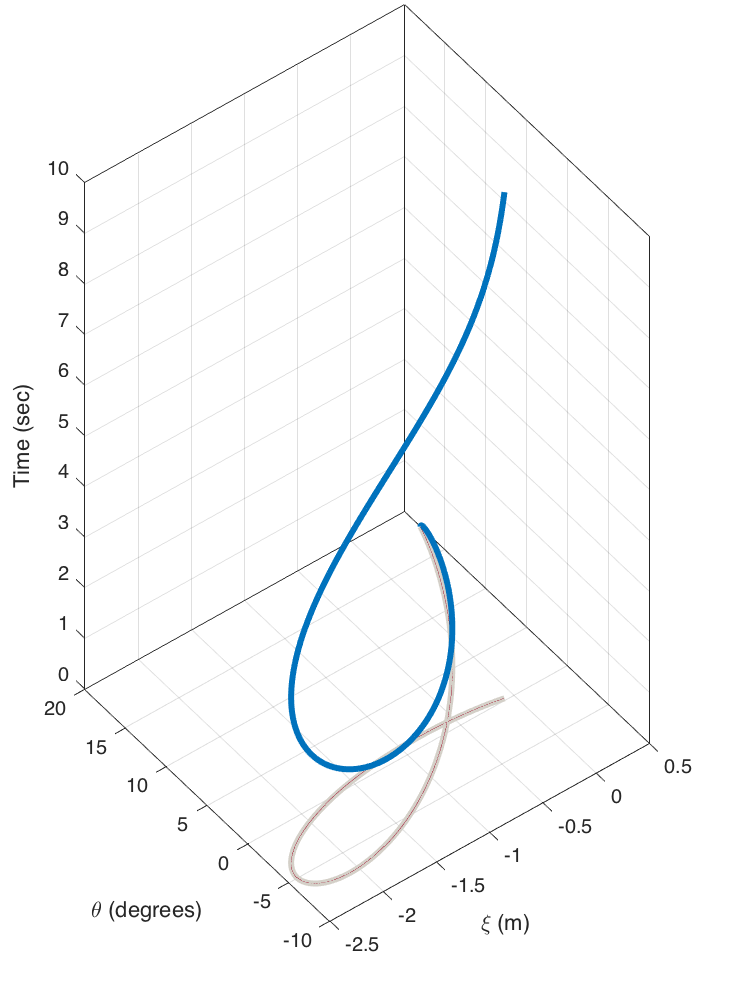}
  \caption{Evolution of ball and beam configurations with time for case 2\\}
  \label{traj1}
\end{figure}

\begin{figure}[H]
\begin{subfigure}{0.5\textwidth}
  \centering
  \includegraphics[scale=0.17]{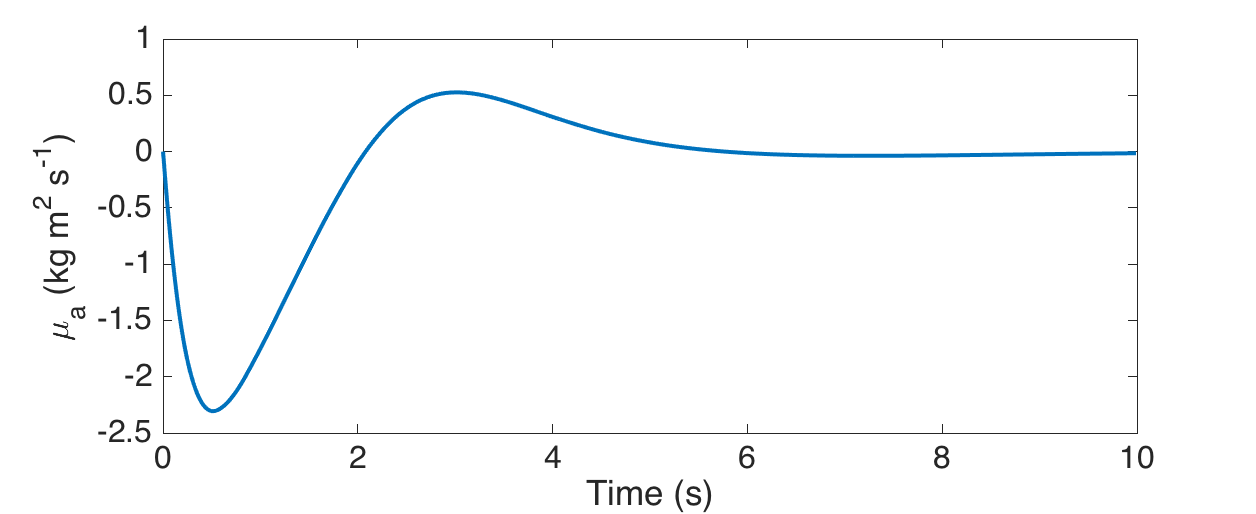}
\end{subfigure}
\begin{subfigure}{0.5\textwidth}
  \centering
  \includegraphics[scale=0.17]{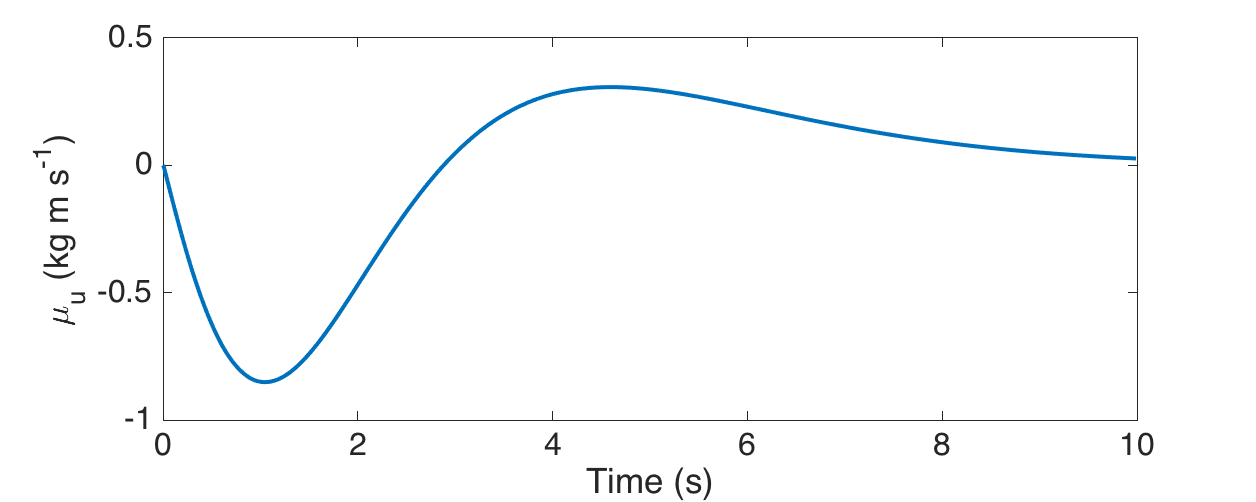}
  \end{subfigure}
  \begin{subfigure}{0.5\textwidth}
  \centering
  \includegraphics[scale=0.17]{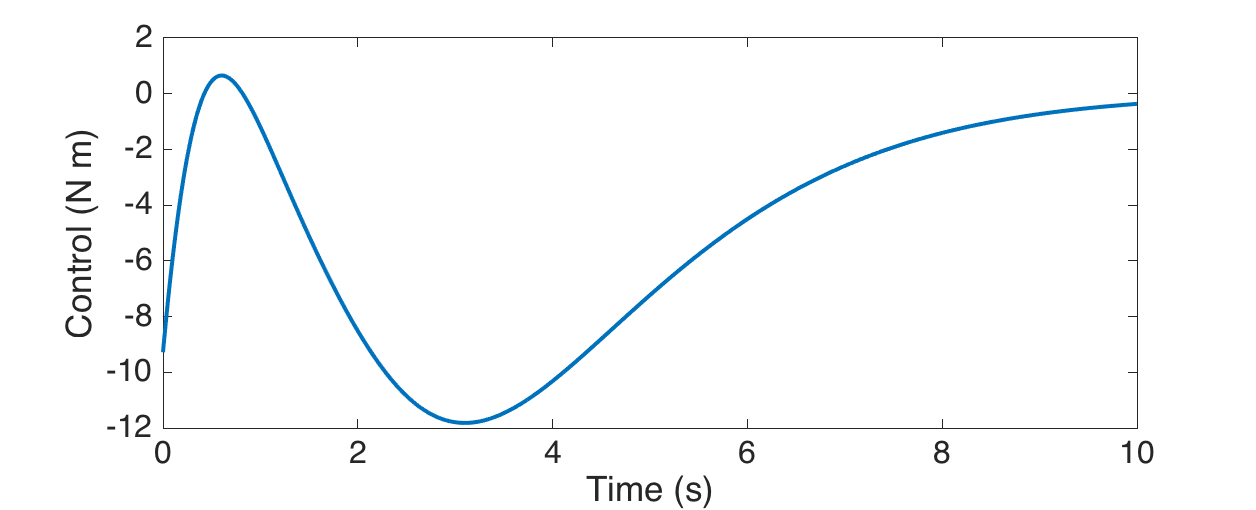}
  \end{subfigure}
  \caption{Momenta and control input versus time for case 2}
\end{figure}


\section{EXAMPLE: Inverted Pendulum on a Cart}
Consider an inverted pendulum with a bob of mass $m_b$ hinged onto a cart of mass $m_c$. The situation is described in the figure below.
\begin{figure}[H]
\centering
\includegraphics[scale=0.4]{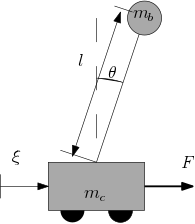}
\caption{Inverted pendulum on a cart}
\end{figure}
$\theta$ is the angle that the pendulum makes with the vertical while $\xi$ is the position of the cart. An external force $F$ acting the cart serves as the control input to the system. The system trajectories evolve over the manifold $Q=S^1\times \mathbb{R}$ with the pendulum constituting the unactuated subsystem, and the cart constituting the actuated subsystem. The Lagrangian of the system is given by
\small
\begin{align*}
&\mathcal{L}\ :\ TQ\rightarrow\mathbb{R}\\
&\mathcal{L} (\xi,\theta,\dot{\xi},\dot{\theta})= \frac{1}{2}(m_b+m_c)\dot{\xi}^2+\frac{1}{2}m_bl^2\dot{\theta}^2-m_bl\dot{\xi}\dot{\theta}\cos\theta-m_bgl\cos\theta
\end{align*}
\normalsize
We now proceed to discretize the system dynamics using the trapezoidal rule. For a time step $h$, the discrete Lagrangian is given by
\small
\begin{align*}
\mathcal{L}_{dk}\ &:\ Q\times Q\rightarrow \mathbb{R}\\
\mathcal{L}_{dk}(\xi_k,\theta_k,\Delta \xi_k, \Delta \theta_k)&=hL(\xi_k,\theta_k, \frac{\Delta \xi_k}{h}, \frac{\Delta \theta_k}{h})\\
&=\frac{1}{2h}(m_b+m_c)\Delta\xi_k^2+\frac{1}{2h}m_bl^2\Delta\theta_k^2\\&-\frac{m_b}{h}l\Delta\xi_k\Delta\theta_k\cos\theta_k-m_bhgl\cos\theta_k
\end{align*}
\normalsize
The discrete-time Euler-Lagrange equations of motion are given by
\small
\begin{subequations}
\begin{align}
&\frac{m_bl^2}{h}(\Delta\theta_{k-1}-\Delta\theta_k)+\frac{m_bl}{h}(\Delta\xi_k\cos\theta_k-\Delta\xi_{k-1}\cos\theta_{k-1})\nonumber\\&+\frac{m_bl}{h}\Delta\xi_k\Delta\theta_k\sin\theta_k+m_bglh\sin\theta_k=0\\
&\frac{m_b+m_c}{h}(\Delta\xi_{k-1}-\Delta\xi_k)+\frac{m_bl}{h}(\Delta\theta_k\cos\theta_k-\Delta\theta_{k-1}\cos\theta_{k-1})\nonumber\\&+\frac{h}{2}(F_k+F_{k-1})=0\\
&\theta_{k}=\theta_{k-1}+\Delta\theta_{k-1}\\
&\xi_{k}=\xi_{k-1}+\Delta \xi_{k-1}
\end{align}
\end{subequations}
\normalsize

The discrete-time Hamilton's equations are given by
\small
\begin{subequations}
\label{heqbb}
\begin{align}
&\mu_{ak}=\frac{m_b+m_c}{h}\Delta\xi_k-\frac{m_bl}{h}\Delta\theta_k\cos\theta_k-\frac{h}{2}F_k\\
&\xi_{k+1}=\xi_{k}+\Delta\xi_{k}\\
&\mu_{ak+1}=\mu_{ak}+\frac{h}{2}F_k+\frac{h}{2}F_{k+1}\\
&\mu_{uk}=-m_bghl\sin\theta_k-\frac{m_bl}{h}\Delta\xi_k\Delta\theta_k\sin\theta_k+\frac{m_bl}{h}(l\Delta\theta_k-\Delta\xi_k\cos\theta_k)\\
&\theta_{k+1}=\theta_{k}+\Delta \theta_{k}\\
&\mu_{uk+1}=\mu_{uk}+\frac{m_bl}{h}\Delta\xi_k\Delta\theta_k\sin\theta_k+m_bglh\sin\theta_k
\end{align}
\end{subequations}
\normalsize
\newpage
Let us find the sequence of controls $\{u_k\}$ which minimizes the cost $\mathcal{J}_d=\sum_{k=0}^{N-1}\frac{1}{2}u^2_k$ and gets the system from $(\theta_0,\xi_0,\mu_{a0},\mu_{u0})=(\theta_0,\xi_0,0,0)$ to $(\theta_N,\xi_N,\mu_{aN},\mu_{uN})=(\theta_f,\xi_f,0,0)$.\\\\
We thus wish to solve the problem 
\begin{align}\label{edocp}
\textrm{min}_{u_k}(\mathcal{J}_d&=\sum_{k=0}^{N-1}\frac{1}{2}u^2_k)\nonumber\\
\textrm{such that }(\theta_0,\xi_0,\mu_{a0},\mu_{u0})&=(\theta_0,\xi_0,0,0)\nonumber\\  (\theta_N,\xi_N,\mu_{aN},\mu_{uN})&=(\theta_f,\xi_f,0,0)\nonumber\\
\textrm{subject to } (\ref{heqbb})
\end{align}
The following table presents the first and second derivatives of the Lagrangian (defined as in (\ref{M}) and (\ref{calM})) in order to obtain the multiplier equations and optimality condition as in (\ref{doce1}) and (\ref{doce2})\\
\begin{center}
\begin{tabular}{ |c|c|c|c| } 
\hline
 $M_{ag}=0$ & $M_{af}=\frac{m_b+m_c}{h}\Delta\xi_k$\\
 & $-\frac{m_bl}{h}\Delta\theta_k\cos\theta_k$\\
 \hline
 $\mathcal{M}^{ag}_{ag}=0$ & $\mathcal{M}^{ag}_{af}=0$\\
$\mathcal{M}^{af}_{ag}=0$ & $\mathcal{M}^{af}_{af}=\frac{m_b+m_c}{h}\lambda$\\
$\mathcal{M}^{ug}_{ag}=0$ & $\mathcal{M}^{ug}_{af}=\frac{m_bl}{h}\Delta\theta_k\sin\theta_k\lambda$\\
$\mathcal{M}^{uf}_{ag}=0$ & $\mathcal{M}^{uf}_{af}=-\frac{m_bl}{h}\cos\theta_k\lambda$\\
\hline
$M_{ug}=\frac{m_bl}{h}\Delta\xi_k\Delta\theta_k\sin\theta_k$ & $M_{uf}=\frac{m_bl^2}{h}\Delta\theta_k$ \\
$+m_bhgl\sin\theta_k$ & $-\frac{m_bl}{h}\Delta\xi_k\cos\theta_k $\\
\hline
$\mathcal{M}^{ag}_{ug}=0$ & $\mathcal{M}^{ag}_{uf}=0$\\
$\mathcal{M}^{af}_{ug}=\frac{m_bl}{h}\Delta\theta_k\sin\theta_k\lambda$ & $\mathcal{M}^{af}_{uf}=-\frac{m_bl}{h}\cos\theta_k\lambda$\\
$\mathcal{M}^{ug}_{ug}=\frac{m_bl}{h}\Delta\xi_k\Delta\theta_k\cos\theta_k\lambda$ & $\mathcal{M}^{ug}_{uf}=\frac{m_bl}{h}\Delta\xi_k\sin\theta_k\lambda$\\
$+m_bhgl\sin\theta_k\lambda$ & \\
$\mathcal{M}^{uf}_{ug}=\frac{m_bl}{h}\Delta\xi_k\sin\theta_k\lambda$ & $\mathcal{M}^{uf}_{uf}=\frac{m_bl^2}{h}\lambda$\\\hline
\end{tabular}
\end{center}
\normalsize
%

The multiplier equations and condition of optimality are therefore given by
\small
\begin{subequations}
\begin{align}
&0=-\lambda^2_{k-1}+\lambda^2_k\\
&-\frac{m_bl}{h}\Delta\theta_k\sin\theta\lambda^1_k-\frac{m_bl}{h}\Delta\xi_k\sin\theta_k\lambda^4_k\nonumber\\&+\frac{m_bl}{h}\Delta\xi_k\Delta\theta_k\cos\theta_k(\lambda^4_k-\lambda^6_k)+m_bhgl\sin\theta_k(\lambda^4_k-\lambda^6_k)\nonumber\\&=-\lambda^5_{k-1}+\lambda^5_k\\
&-\frac{m_b+m_c}{h}\lambda^1_k+\frac{m_bl}{h}\Delta\theta_k\sin\theta_k(\lambda^4_k-\lambda^6_k)+\frac{m_bl}{h}\cos\theta_k\lambda^4_k\nonumber\\&=\lambda^2_k\\
&\frac{m_bl}{h}\cos\theta_k\lambda^1_k+\frac{m_bl}{h}\Delta\xi_k\sin\theta_k(\lambda^4_k-\lambda^6_k)-\frac{m_bl^2}{h}\lambda^4_k=\lambda^5_k\\
&\lambda^1_k-\lambda^3_k=-\lambda^3_{k-1}\\
&\lambda^4_k-\lambda^6_k=-\lambda^6_{k-1}\\
&F_k+\frac{h}{2}\lambda^1_k-\frac{h}{2}\lambda^3_k=\frac{h}{2}\lambda^3_{k-1}\hspace{1cm} \forall k=1,2,\dots, N-1\\
&F_0+\frac{h}{2}\lambda^1_0-\frac{h}{2}\lambda^3_0=0
\end{align}
\end{subequations}
\normalsize

The solutions to these equations are  also solved using the multiple shooting method described in \cite{karmvir}.\\
\subsection*{Simulation Results}
As in the previous example, we obtain the solution numerically on a PC by implementing the above algorithm with the following parameters
\begin{center}
 \begin{tabular}{||c c||} 
 \hline
 Parameter & Value  \\ [0.5ex] 
 \hline\hline
 $m_c$ & $0.5\ kg$ \\ 
 \hline
 $m_b$ & $0.1\ kg$ \\
 \hline
  $l$ & $0.1\ m$ \\
 \hline
 $g$ & $9.8\ m\ s^{-2}$ \\
 \hline
 $h$ & $0.01\ s$ \\
 \hline
 N & 1000\\ [1ex] 
 \hline
\end{tabular}
\end{center}
We present our results for two sets of boundary conditions 
\subsubsection*{\textbf{Case 1}}
The initial and terminal conditions are set as 
\begin{center}
 \begin{tabular}{||c c c c||} 
 \hline\hline
 $\theta_0= $& $60^{\circ}$ &  $\theta_N =$& 0 \\ 
 \hline
 $\mu_{a0}=$ & $0\ kg\ m^2\ s^{-1}$ &  $\mu_{aN}=$ & $0\ kg\ m^2\ s^{-1}$ \\
 \hline
 $\xi_0 =$& $2\ m$ &  $\xi_N=$ & $0\ m$ \\
 \hline
 $\mu_{u0}=$ & $0\ kg\ m\ s^{-1}$ & $\mu_{uN}=$ & $0\ kg\ m\ s^{-1}$ \\[1ex] 
 \hline\hline
\end{tabular}
\end{center}

\begin{figure}[H]
  \centering
  \includegraphics[scale=0.22]{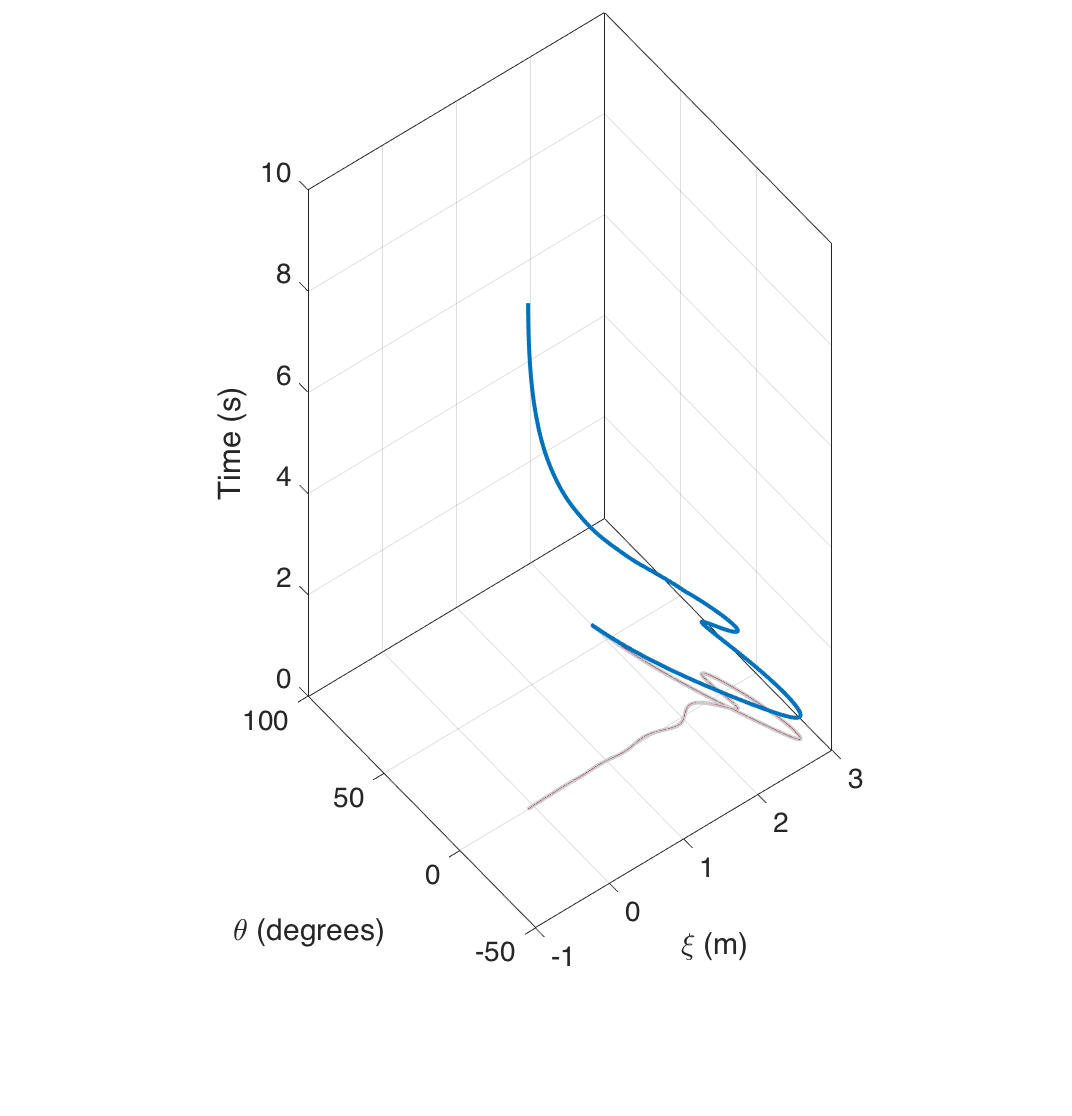}
  \caption{Evolution of pendulum and cart configurations with time for case 1}
  \label{traj1}
\end{figure}

\begin{figure}[H]
\begin{subfigure}{0.5\textwidth}
  \centering
  \includegraphics[scale=0.13]{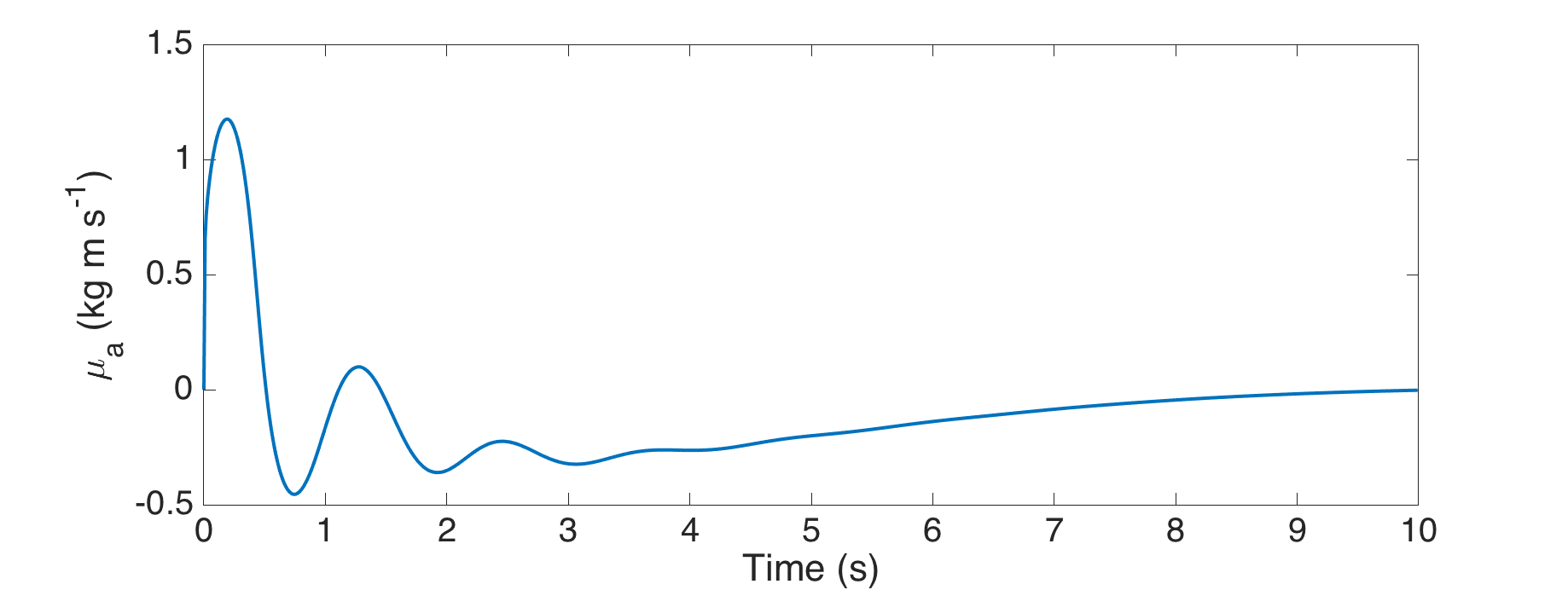}
\end{subfigure}
\begin{subfigure}{0.5\textwidth}
  \centering
  \includegraphics[scale=0.13]{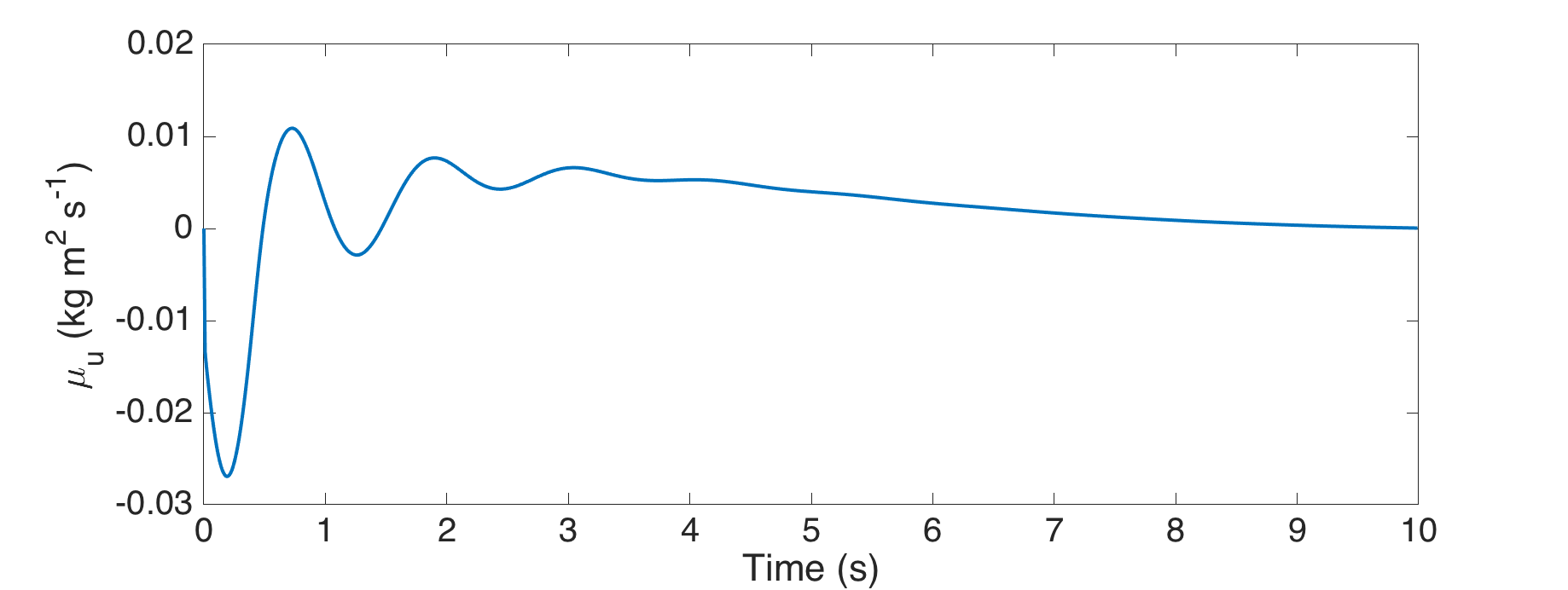}
  \end{subfigure}
  \begin{subfigure}{0.5\textwidth}
  \centering
  \includegraphics[scale=0.13]{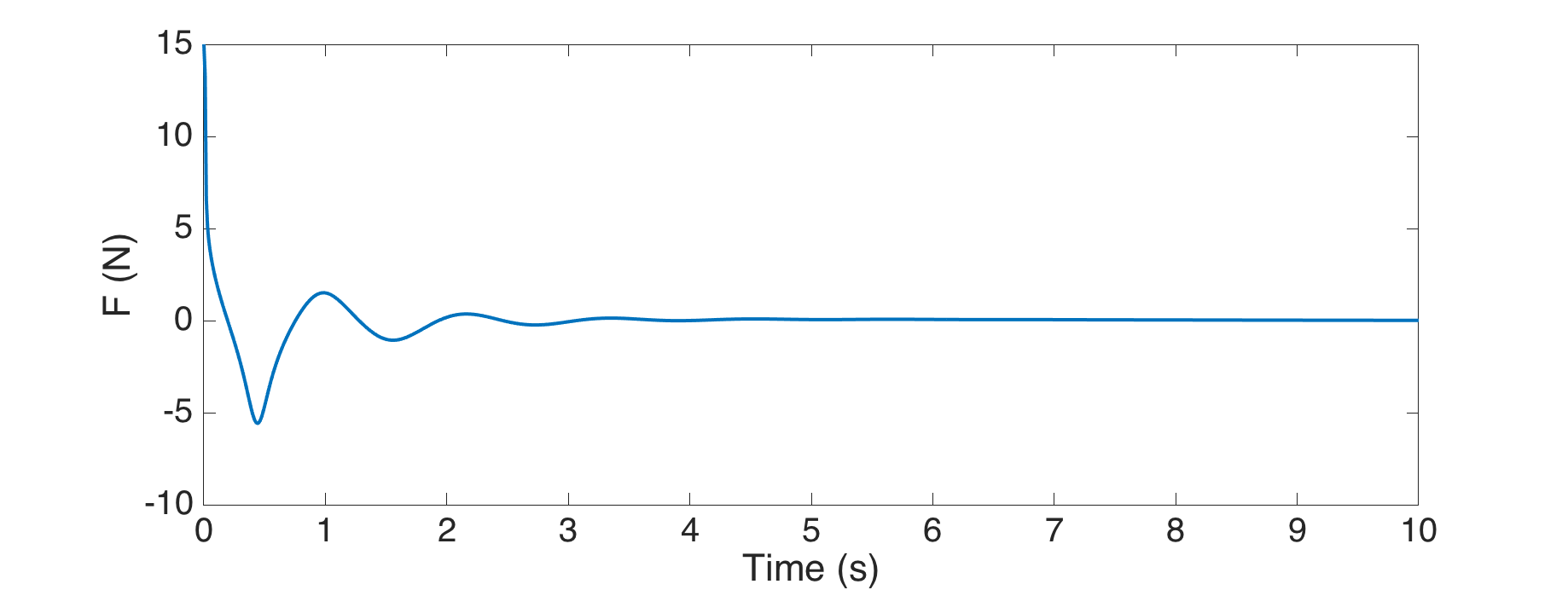}
  \end{subfigure}
  \caption{Momenta and control input versus time for case 1}
\end{figure}
We see that the numerical solution obtained respects the boundary conditions satisfactorily, stabilizing both, the ball and the beam.
\subsubsection*{\textbf{Case 2}}
The initial and terminal conditions are set as 
\begin{center}
 \begin{tabular}{||c c c c||} 
 \hline\hline
 $\theta_0= $& $-45^{\circ}$ &  $\theta_N =$& 0 \\ 
 \hline
 $\mu_{a0}=$ & $0\ kg\ m^2\ s^{-1}$ &  $\mu_{aN}=$ & $0\ kg\ m^2\ s^{-1}$ \\
 \hline
 $\xi_0 =$& $2\ m$ &  $\xi_N=$ & $0\ m$ \\
 \hline
 $\mu_{u0}=$ & $0\ kg\ m\ s^{-1}$ & $\mu_{uN}=$ & $0\ kg\ m\ s^{-1}$ \\[1ex] 
 \hline\hline
\end{tabular}
\end{center}

\begin{figure}[H]
  \centering
  \includegraphics[scale=0.19]{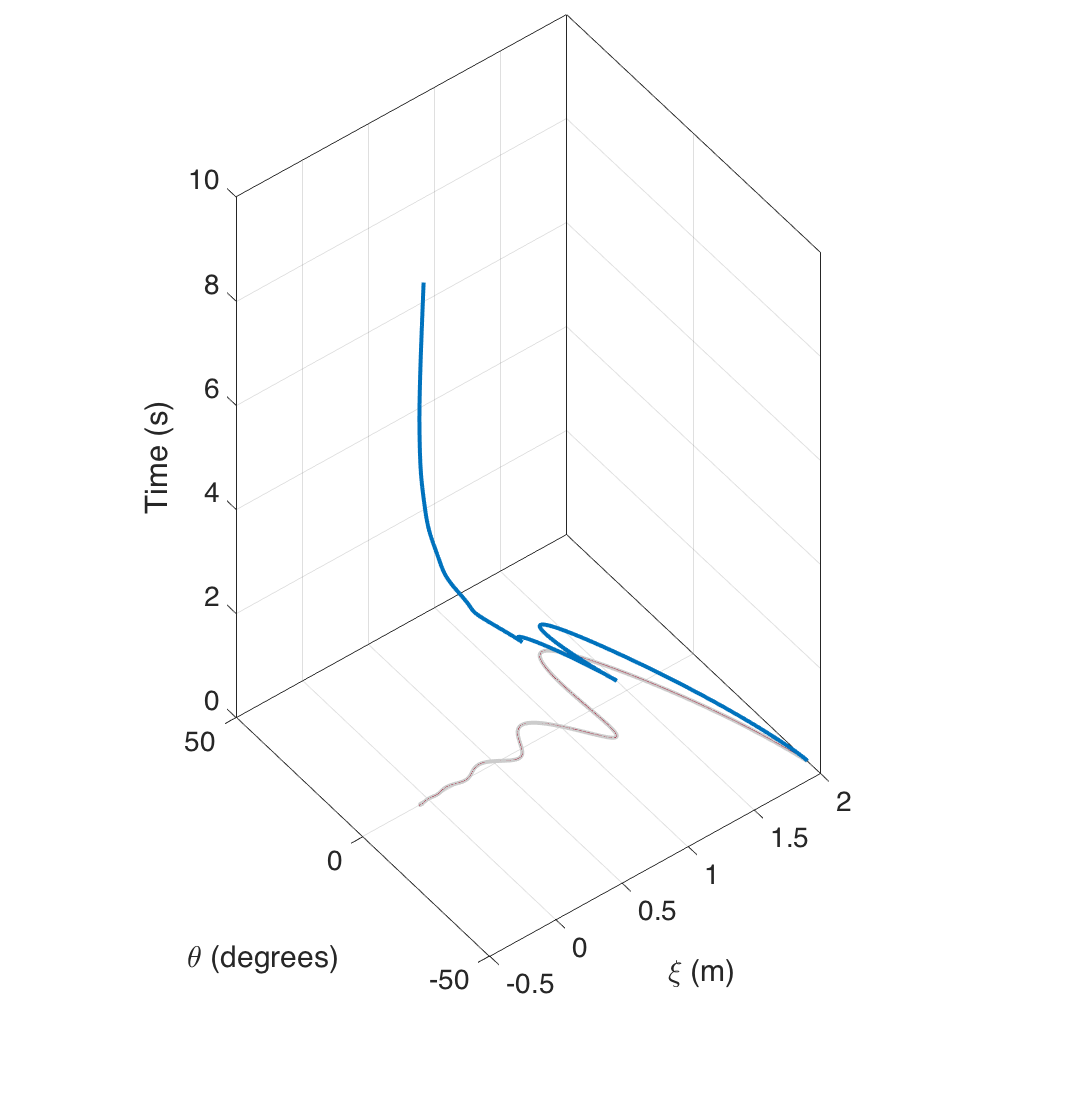}
  \caption{Evolution of pendulum and cart configurations with time for case 2\\}
  \label{traj1}
\end{figure}

\begin{figure}[H]
\begin{subfigure}{0.5\textwidth}
  \centering
  \includegraphics[scale=0.13]{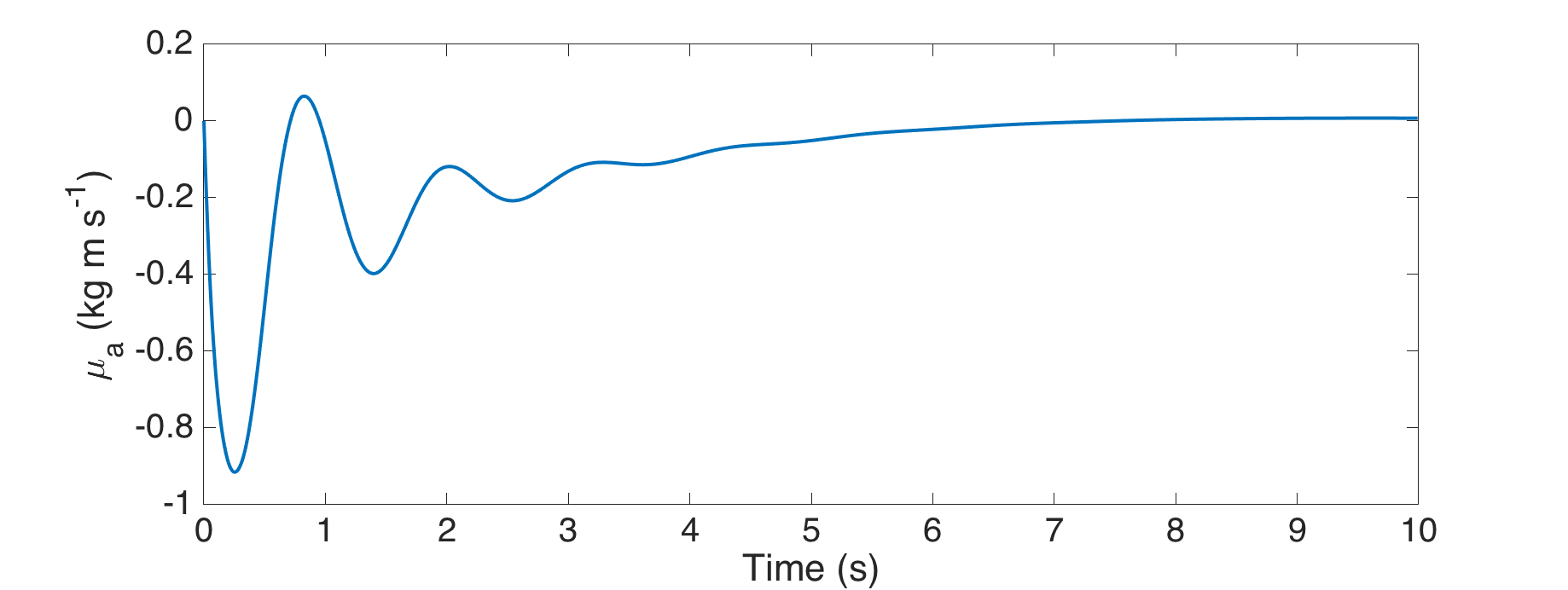}
\end{subfigure}
\begin{subfigure}{0.5\textwidth}
  \centering
  \includegraphics[scale=0.13]{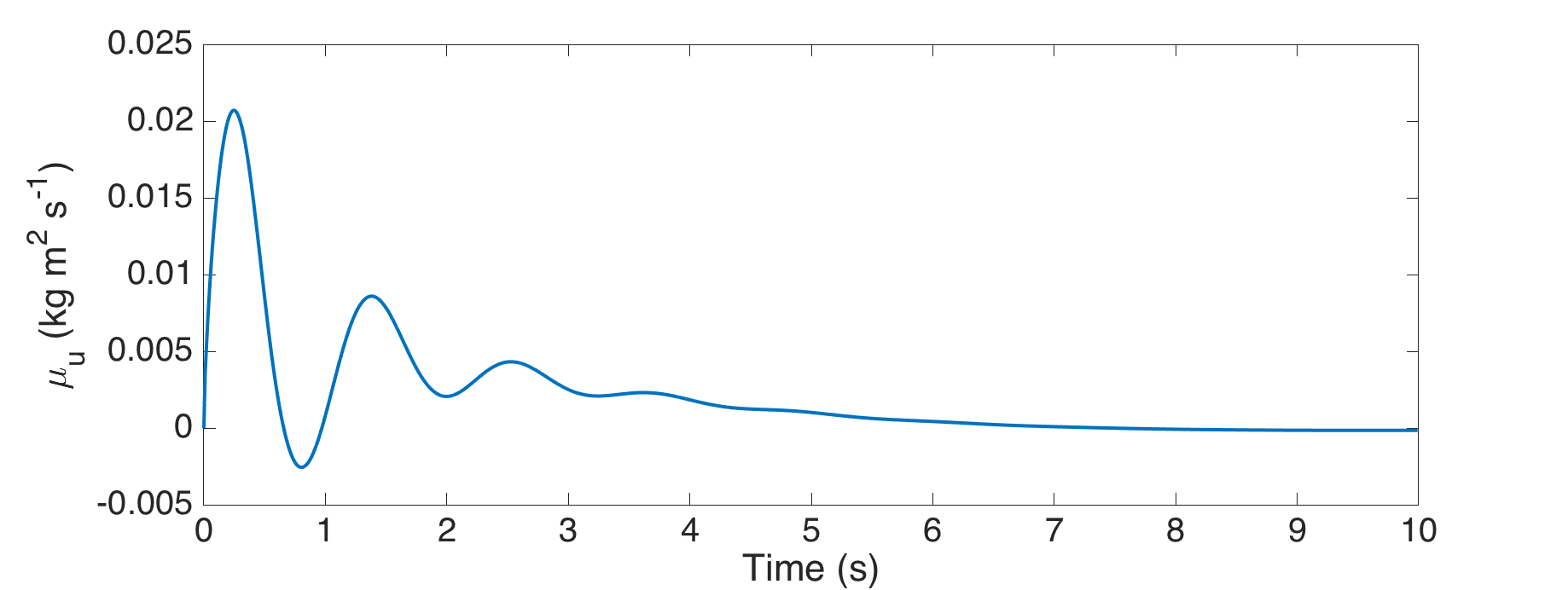}
  \end{subfigure}
  \begin{subfigure}{0.5\textwidth}
  \centering
  \includegraphics[scale=0.13]{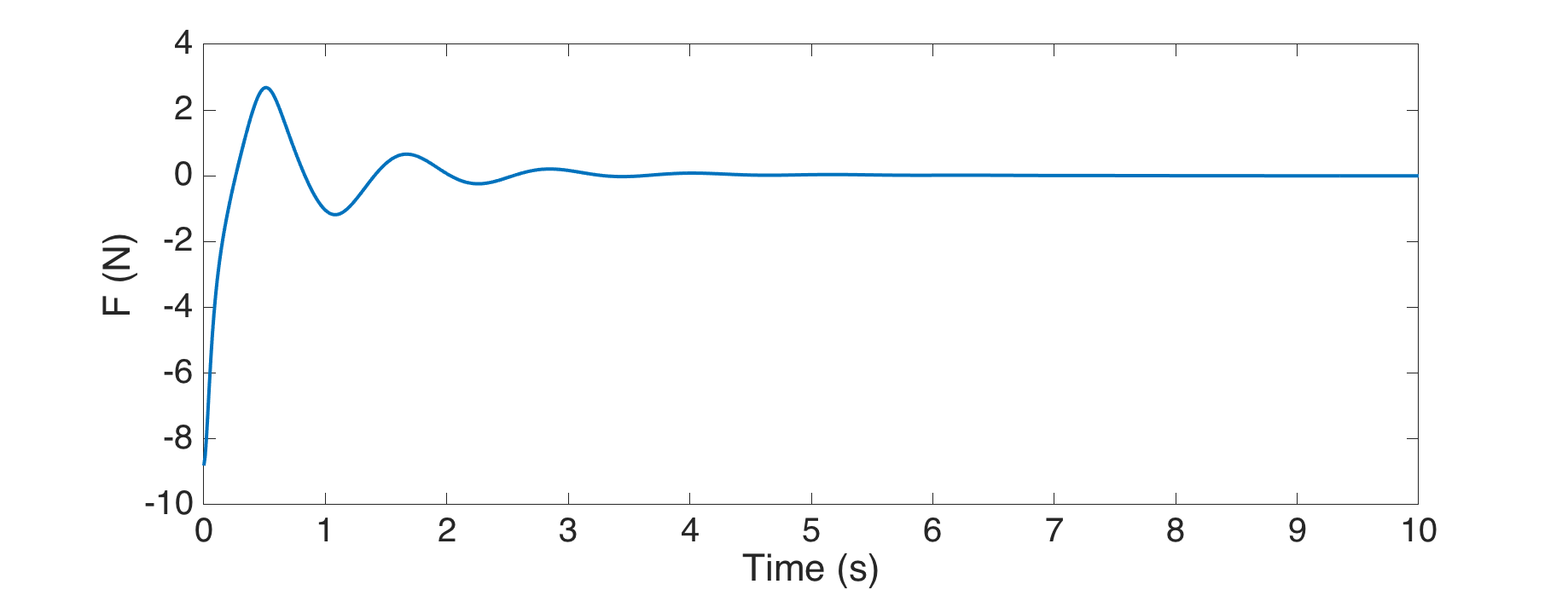}
  \end{subfigure}
  \caption{Momenta and control input versus time for case 2}
\end{figure}


\section{Conclusion}
In this article, we developed a variational integrator for interconnected mechanical systems evolving on a product of matrix Lie groups. A discrete optimal control problem was formulated for the considered class of systems and subsequently solved using calculus of variations to obtain necessary conditions describing optimal trajectories. The proposed approach is demonstrated on a benchmark underactuated system with satisfactory results. The discrete optimal control problem solved here is that of finding an optimal trajectory, given fixed endpoints. An extension of this work would to be solve a more general class of problems. Moreover, the conditions of optimality obtained in this work are merely necessary conditions that an optimal trajectory should possess. The existence of the same is not guaranteed. A starting attempt to answer this question would be to analyse the controllability of interconnected mechanical systems considered here.
\bibliographystyle{IEEEtran}
\bibliography{IEEEexample}


\section{APPENDIX}

\subsection*{Obtaining the necessary conditions}

With the introduction of Lagrange multipliers, $\mu_{ak}, \mu_{uk}, g_{uk}, g_{ak}, f_{uk}, f_{ak}$ are varied independently. Using one-parameter subgroups on their respective Lie groups, the variations of the last four terms are given by
\begin{align*}
&\delta g_{uk}=g_{uk}\eta_{uk} \quad \delta g_{ak}=g_{ak}\eta_{ak}\\
&\delta f_{uk}=f_{uk}\chi_{uk} \quad \delta f_{ak}=f_{ak}\chi_{ak}
\end{align*}

for some $\eta_{uk}, \chi_{uk}\ \in \mathfrak{g}_u$ and $\eta_{ak}, \chi_{ak} \in \mathfrak{g}_a$. We'll require the following results to calculate the variations of the terms in (\ref{vj1}).\\
For $g\in G$, the adjoint operator $\textrm{Ad}_g:\mathfrak{g}\rightarrow\mathfrak{g}$ is the tangent lift of the inner automorphism
\begin{align}
\textrm{Ad}_g\xi=T_{g^{-1}}L_{g}\cdot T_eR_{g^{-1}}\cdot\xi
\end{align}
The derivative of $\textrm{Ad}_g\xi$ with respect to $g$ at $e$ in the direction $\eta$ gives us the ad operator $\textrm{ad}_\eta\xi=[\eta,\xi]$
\begin{align}
\left.\dfrac{d}{d\epsilon}\right|_{\epsilon=0}\textrm{Ad}_{\textrm{exp}\ \epsilon\eta}\xi=[\eta,\xi]
\end{align}
\begin{prop}
The derivatives of the $Ad$ map are given by
\small
\begin{subequations}
\label{add}
\begin{align}
 &\left.\dfrac{d}{d\epsilon}\right|_{\epsilon=0}\textrm{Ad}_{g\exp{\ \epsilon\eta}}\xi=\textrm{Ad}_g[\eta,\xi]=[\textrm{Ad}_g\eta,\textrm{Ad}_g\xi]\\
&\left.\dfrac{d}{d\epsilon}\right|_{\epsilon=0}\textrm{Ad}_{(g\exp{\ \epsilon\eta})^{-1}}\xi=[\textrm{Ad}_{g^{-1}}\xi,\eta]\\
&\left.\dfrac{d}{d\epsilon}\right|_{\epsilon=0}\textrm{Ad}^*_{g\exp{\ \epsilon\eta}}\alpha=\textrm{Ad}^*_g(\textrm{ad}^*_{\textrm{Ad}_g\eta}\alpha)\\
&\left.\dfrac{d}{d\epsilon}\right|_{\epsilon=0}\textrm{Ad}^*_{(g\exp{\ \epsilon\eta})^{-1}}\alpha=-\textrm{Ad}^*_{g^{-1}}(\textrm{ad}^*_{\eta}\alpha)
\end{align}
\end{subequations}
\normalsize
\end{prop}
\textbf{\\\textit{Proof}:}
\small
\begin{align*}
 \left.\dfrac{d}{d\epsilon}\right|_{\epsilon=0}\textrm{Ad}_{g\exp{\ \epsilon\eta}}\xi&=\left.\dfrac{d}{d\epsilon}\right|_{\epsilon=0}\textrm{Ad}_g\circ \textrm{Ad}_{\exp{\ \epsilon\eta}}\xi\\
 &=\textrm{Ad}_g[\eta,\xi]\\
 &=[\textrm{Ad}_g\eta,\textrm{Ad}_g\xi]
\end{align*}
\begin{align*}
\left.\dfrac{d}{d\epsilon}\right|_{\epsilon=0}\textrm{Ad}_{(g\exp{\ \epsilon\eta})^{-1}}\xi&=\left.\dfrac{d}{d\epsilon}\right|_{\epsilon=0}\textrm{Ad}_{\exp{-\epsilon\eta}}\circ\textrm{Ad}_{g^{-1}}\xi\\
&=-[\eta, \textrm{Ad}_{g^{-1}}\xi]\\
&=[\textrm{Ad}_{g^{-1}}\xi,\eta]
\end{align*}
\begin{align*}
\langle \left.\dfrac{d}{d\epsilon}\right|_{\epsilon=0}\textrm{Ad}^*_{g\exp{\ \epsilon\eta}}\alpha, \xi\rangle&=\langle\alpha, \left.\dfrac{d}{d\epsilon}\right|_{\epsilon=0}\textrm{Ad}_{g\exp{\ \epsilon\eta}}\xi\rangle\\
&=\langle\alpha, [\textrm{Ad}_g\eta,\textrm{Ad}_g\xi]\rangle\\
&=\langle\alpha, \textrm{ad}_{\textrm{Ad}_g\eta}(\textrm{Ad}_g\xi)\rangle\\
&=\langle\textrm{Ad}^*_g(\textrm{ad}^*_{\textrm{Ad}_g\eta}\alpha),\xi\rangle
\end{align*}
\begin{align*}
\langle \left.\dfrac{d}{d\epsilon}\right|_{\epsilon=0}\textrm{Ad}^*_{(g\exp{\ \epsilon\eta})^{-1}}\alpha, \xi\rangle&=\langle\alpha, \left.\dfrac{d}{d\epsilon}\right|_{\epsilon=0}\textrm{Ad}_{(g\exp{\ \epsilon\eta})^{-1}}\xi\rangle\\
&=\langle\alpha, -[\eta,\textrm{Ad}_{g^{-1}}\xi]\rangle\\
&=\langle\alpha, -\textrm{ad}_{\eta}(\textrm{Ad}_{g^{-1}}\xi)\rangle\\
&=\langle-\textrm{Ad}^*_{g^{-1}}(\textrm{ad}^*_{\eta}\alpha),\xi\rangle
\end{align*}
\normalsize
\hfill$\blacksquare$\\\\
Now we proceed to obtain the variations of the terms in (\ref{vj1})\\\\

\footnotesize
\begin{align*}
\delta \mathcal{J}_{d0}&=D_{g_{ak}}\phi_d(g_k,f_k,u_k)\cdot\delta g_{ak}+D_{g_{uk}}\phi_d(g_k,f_k,u_k)\cdot\delta g_{uk}\\&+D_{f_{ak}}\phi_d(g_k,f_k,u_k)\cdot\delta f_{ak}+D_{f_{uk}}\phi_d(g_k,f_k,u_k)\cdot\delta f_{uk}\\&+D_{u_{k}}\phi_d(g_k,f_k,u_k)\cdot\delta u_{k}\\
&=\langle T^*_eL_{g_{ak}}\cdot D_{g_{ak}}\phi_d, \eta_{ak}\rangle+\langle T^*_eL_{g_{uk}}\cdot D_{g_{uk}}\phi_d, \eta_{uk}\rangle\\&+\langle T^*_eL_{f_{ak}}\cdot D_{f_{ak}}\phi_d, \chi_{ak}\rangle+\langle T^*_eL_{f_{uk}}\cdot D_{f_{uk}}\phi_d, \chi_{uk}\rangle\\
&+\langle\delta u_k, D_{u_{k}}\phi_d(g_k,f_k,u_k)\rangle
\end{align*}

\begin{align*}
\delta \mathcal{J}_{d1}&=\delta(\langle\mu_{ak}-(-M_{ag}+\textrm{Ad}^*_{f^{-1}_{ak}}\cdot(M_{af})-\frac{1}{2}hu_k), \lambda^1_k\rangle)\\
&=\langle\delta\mu_{ak}+\frac{h}{2}\delta u_k+D_{g_{ak}}M_{ag}\cdot\delta g_{ak}+D_{g_{uk}}M_{ag}\cdot\delta g_{uk}+D_{f_{ak}}M_{ag}\cdot\delta f_{ak}\\&+D_{f_{uk}}M_{ag}\cdot\delta f_{uk}, \lambda^1_k\rangle+\langle-\textrm{Ad}^*_{f^{-1}_{ak}}(D_{g_{ak}}M_{af}\cdot\delta g_{ak}+D_{g_{uk}}M_{af}\cdot\delta g_{uk}\\&+D_{f_{ak}}M_{af}\cdot\delta f_{ak}+D_{f_{uk}}M_{af}\cdot\delta f_{uk})+\textrm{Ad}^*_{f_{ak}^{-1}}(\textrm{ad}^*_{\chi_{ak}}M_{af}),\lambda^1_k\rangle\\
&=\langle \mathcal{M}^{ag}_{ag}(\lambda^1_k)-\mathcal{M}^{ag}_{af}(\textrm{Ad}_{f^{-1}_{ak}}\lambda^1_k),\eta_{ak}\rangle+\langle \mathcal{M}^{ug}_{ag}(\lambda^1_k)-\mathcal{M}^{ug}_{af}(\textrm{Ad}_{f^{-1}_{ak}}\lambda^1_k),\eta_{uk}\rangle\\&+\langle \mathcal{M}^{af}_{ag}(\lambda^1_k)-\mathcal{M}^{af}_{af}(\textrm{Ad}_{f^{-1}_{ak}}\lambda^1_k),\chi_{ak}\rangle+\langle \mathcal{M}^{uf}_{ag}(\lambda^1_k)-\mathcal{M}^{uf}_{af}(\textrm{Ad}_{f^{-1}_{ak}}\lambda^1_k),\chi_{uk}\rangle\\
&+\langle -\textrm{ad}^*_{\textrm{Ad}_{f_{ak}^{-1}}\lambda^1_k}(M_{af}),\chi_{ak}\rangle+\langle\delta\mu_{ak}+\frac{h}{2}\delta u_k,\lambda^1_k\rangle
\end{align*}

\begin{align*}
\delta \mathcal{J}_{d3}&=\delta(\langle \mu_{ak+1}-(\textrm{Ad}^*_{f_{ak}}(\mu_{ak}+M_{ag}+\frac{1}{2}hu_k)+\frac{1}{2}hu_{k+1}), \lambda^3_k\rangle)\\
&=\langle\delta\mu_{ak+1}-\frac{h}{2}\delta u_{k+1}-\textrm{Ad}^*_{f_{ak}}(\delta\mu_{ak}+\frac{h}{2}\delta u_k+\textrm{ad}^*_{\textrm{Ad}_{f_{ak}}\chi_{ak}}M_{ag}),\lambda^3_k\rangle\\
&+\langle-\textrm{Ad}^*_{f_{ak}}(D_{g_{ak}}M_{ag}\cdot\delta g_{ak}+D_{g_{uk}}M_{ag}\cdot\delta g_{uk}+D_{f_{ak}}M_{ag}\cdot\delta f_{ak}\\&+D_{f_{uk}}M_{ag}\cdot\delta f_{uk}),\lambda^3_k\rangle\\
&=\langle\delta\mu_{ak+1},\lambda^3_k\rangle-\langle\delta\mu_{ak},\textrm{Ad}_{f_{ak}}\lambda^3_k\rangle-\langle\delta u_k, \frac{h}{2}\textrm{Ad}_{f_{ak}}\lambda^3_k\rangle-\langle\delta u_{k+1}, \frac{h}{2}\lambda^3_k\rangle
\\&+\langle \textrm{Ad}^*_{f_{ak}}\textrm{ad}^*_{\textrm{Ad}_{f_{ak}}\lambda^3_k}(\mu_{ak}+M_{ag}+\frac{1}{2}hu_k),\chi_{ak}\rangle
-\langle\mathcal{M}^{ag}_{ag}(\textrm{Ad}_{f_{ak}}\lambda^3_k),\eta_{ak}\rangle\\&-\langle \mathcal{M}^{ug}_{ag}(\textrm{Ad}_{f_{ak}} \lambda^3_k),\eta_{uk}\rangle-\langle \mathcal{M}^{af}_{ag}(\textrm{Ad}_{f_{ak}}\lambda^3_k),\chi_{ak}\rangle-\langle \mathcal{M}^{uf}_{ag}(\textrm{Ad}_{f_{ak}}\lambda^3_k),\chi_{uk}\rangle
\end{align*}

\begin{align*}
\delta \mathcal{J}_{d4}&=\delta(\langle\mu_{uk}-(-M_{ug}+\textrm{Ad}^*_{f^{-1}_{uk}}\cdot(M_{uf})), \lambda^4_k\rangle)\\
&=\langle\delta\mu_{uk}+D_{g_{ak}}M_{ug}\cdot\delta g_{ak}+D_{g_{uk}}M_{ug}\cdot\delta g_{uk}+D_{f_{ak}}M_{ug}\cdot\delta f_{ak}\\&+D_{f_{uk}}M_{ug}\cdot\delta f_{uk}, \lambda^4_k\rangle+\langle-\textrm{Ad}^*_{f^{-1}_{uk}}(D_{g_{ak}}M_{uf}\cdot\delta g_{ak}+D_{g_{uk}}M_{uf}\cdot\delta g_{uk}\\&+D_{f_{ak}}M_{uf}\cdot\delta f_{ak}+D_{f_{uk}}M_{uf}\cdot\delta f_{uk})+\textrm{Ad}^*_{f_{uk}^{-1}}(\textrm{ad}^*_{\chi_{uk}}M_{uf}),\lambda^4_k\rangle\\
&=\langle \mathcal{M}^{ag}_{ug}(\lambda^4_k)-\mathcal{M}^{ag}_{uf}(\textrm{Ad}_{f^{-1}_{uk}}\lambda^4_k),\eta_{ak}\rangle\\&+\langle \mathcal{M}^{ug}_{ug}(\lambda^4_k)-\mathcal{M}^{ug}_{uf}(\textrm{Ad}_{f^{-1}_{uk}}\lambda^4_k),\eta_{uk}\rangle\\&+\langle \mathcal{M}^{af}_{ug}(\lambda^4_k)-\mathcal{M}^{af}_{uf}(\textrm{Ad}_{f^{-1}_{uk}}\lambda^4_k),\chi_{ak}\rangle\\&+\langle \mathcal{M}^{uf}_{ug}(\lambda^4_k)-\mathcal{M}^{uf}_{uf}(\textrm{Ad}_{f^{-1}_{uk}}\lambda^4_k),\chi_{uk}\rangle+\langle -\textrm{ad}^*_{\textrm{Ad}_{f_{uk}^{-1}}\lambda^4_k}(M_{uf}),\chi_{uk}\rangle\\&+\langle\delta\mu_{uk},\lambda^4_k\rangle\\
\end{align*}

\begin{align*}
\delta \mathcal{J}_{d6}&=\delta(\langle \mu_{uk+1}-(\textrm{Ad}^*_{f_{uk}}(\mu_{uk}+M_{ug})), \lambda^6_k\rangle)\\
&=\langle\delta\mu_{uk+1}-\textrm{Ad}^*_{f_{uk}}(\delta\mu_{uk}+D_{g_{ak}}M_{ug}\cdot\delta g_{ak}+D_{g_{uk}}M_{ug}\cdot\delta g_{uk}\\&+D_{f_{ak}}M_{ug}\cdot\delta f_{ak}+D_{f_{uk}}M_{ug}\cdot\delta f_{uk})\\&-\textrm{Ad}^*_{f_{uk}}\textrm{ad}^*_{\textrm{Ad}_{f_{uk}}\chi_{uk}}(\mu_{uk}+M_{ug}),\lambda^6_k\rangle\\
&=\langle\delta\mu_{uk+1},\lambda^6_k\rangle+\langle\delta\mu_{uk},-\textrm{Ad}_{f_{uk}}\lambda^6_k\rangle\\&+\langle \textrm{Ad}^*_{f_{uk}}\textrm{ad}^*_{\textrm{Ad}_{f_{uk}}\lambda^6_k}(\mu_{uk}+M_{ug}),\chi_{uk}\rangle
-\langle\mathcal{M}^{ag}_{ug}(\textrm{Ad}_{f_{uk}}\lambda^6_k),\eta_{ak}\rangle\\&-\langle \mathcal{M}^{ug}_{ug}(\textrm{Ad}_{f_{uk}} \lambda^6_k),\eta_{uk}\rangle-\langle \mathcal{M}^{af}_{ug}(\textrm{Ad}_{f_{uk}}\lambda^6_k),\chi_{ak}\rangle\\&-\langle \mathcal{M}^{uf}_{ug}(\textrm{Ad}_{f_{uk}}\lambda^6_k),\chi_{uk}\rangle\\
\end{align*}
\normalsize
Now we compute the variations of the log terms using the BCH formula.

\footnotesize
\begin{align*}
\textrm{log}(g^{\epsilon -1}_{uk}g^{\epsilon}_{uk+1})&=\textrm{log}(\exp(-\epsilon\eta_{uk})g^{-1}_{uk}g_{uk+1}\exp(\epsilon\eta_{uk+1}))\\
&=\log(\exp(-\epsilon\eta_{uk})f_{uk}\exp(\epsilon\eta_{uk+1}))\\
&=\log(f^{-1}_{uk}\exp(-\epsilon\textrm{Ad}_{f^{-1}_{uk}}\eta_{uk})\exp(\epsilon\eta_{uk+1}))
\end{align*}
\normalsize

Setting \footnotesize$$\log(f^{-1}_{uk})=X_{uk}$$$$\log(\exp(-\epsilon\textrm{Ad}_{f^{-1}_{uk}}\eta_{uk})\exp(\epsilon\eta_{uk+1}))=Y_{uk},$$\normalsize
the equation can now be written as

\footnotesize
\begin{align*}
\textrm{log}(g^{\epsilon -1}_{uk}g^{\epsilon}_{uk+1})&=X_{uk}+\dfrac{\textrm{ad}_{X_{uk}}\exp(\textrm{ad}_{X_{uk}})}{\exp(\textrm{ad}_{X_{uk}})-1}Y_{uk}+O(Y^2_{uk})
\end{align*}
\normalsize

Using the definition of a variation, we obtain the following expression. Formula (\ref{bch1}) is used for expanding $\textrm{log}(g^{\epsilon -1}_{uk}g^{\epsilon}_{uk+1})$ while formula (\ref{bch2}) is used to expand $Y_{uk}$.

\footnotesize
\begin{align*}
\delta\textrm{log}(g^{\epsilon -1}_{uk}g^{\epsilon}_{uk+1})&=\left.\dfrac{d}{d\epsilon}\right|_{\epsilon=0}\textrm{log}(g^{\epsilon -1}_{uk}g^{\epsilon}_{uk+1})\\
&= \left.\dfrac{d}{d\epsilon}\right|_{\epsilon=0}X_{uk}+\dfrac{\textrm{ad}_{X_{uk}}\exp(\textrm{ad}_{X_{uk}})}{\exp(\textrm{ad}_{X_{uk}})-1}Y_{uk}+O(Y^2_{uk})\\
&= \left.\dfrac{d}{d\epsilon}\right|_{\epsilon=0}X_{uk}+\dfrac{\textrm{ad}_{X_{uk}}\exp(\textrm{ad}_{X_{uk}})}{\exp(\textrm{ad}_{X_{uk}})-1}(-\epsilon\textrm{Ad}_{f^{-1}_{uk}}\eta_{uk}\\&+\epsilon\eta_{uk+1}+O(\epsilon^2))+O(\epsilon^4)\\
&=\dfrac{\textrm{ad}_{X_{uk}}\exp(\textrm{ad}_{X_{uk}})}{\exp(\textrm{ad}_{X_{uk}})-1}(\eta_{uk+1}-\textrm{Ad}_{f^{-1}_{uk}}\eta_{uk})
\end{align*}
\normalsize
Similarly,
\footnotesize
\begin{align*}
\delta(-\log(f^{\epsilon}_{uk}))&=\left.\dfrac{d}{d\epsilon}\right|_{\epsilon=0}\log(\exp(-\epsilon\chi_{uk})f^{-1}_{uk})\\
&=\dfrac{\textrm{ad}_{X_{uk}}\exp(\textrm{ad}_{X_{uk}})}{\exp(\textrm{ad}_{X_{uk}})-1}(-\chi_{uk})
\end{align*}
\normalsize
This helps us obtain the variations of $\mathcal{J}_{d2}$ and $\mathcal{J}_{d5}$ as follows
\footnotesize
\begin{align*}
\delta\mathcal{J}_{d2}&=\langle\lambda^2_k,\dfrac{\textrm{ad}_{X_{ak}}\exp(\textrm{ad}_{X_{ak}})}{\exp(\textrm{ad}_{X_{ak}})-1}(\eta_{ak+1}-\textrm{Ad}_{f^{-1}_{ak}}\eta_{ak}-\chi_{ak})\rangle\\
&\textrm{Setting }\lambda^2_k\equiv\left(\dfrac{\textrm{ad}_{X_{ak}}\exp(\textrm{ad}_{X_{ak}})}{\exp(\textrm{ad}_{X_{ak}})-1}\right)^*\lambda^2_k,\\
&=\langle \lambda^2_k, \eta_{ak+1}-\textrm{Ad}_{f^{-1}_{ak}}\eta_{ak}-\chi_{ak}\rangle\\
&=\langle \lambda^2_k, \eta_{ak+1}\rangle+\langle-\textrm{Ad}^*_{f^{-1}_{ak}}\lambda^2_k,\eta_{ak}\rangle+\langle-\lambda^2_k,\chi_{ak}\rangle
\end{align*}
\begin{align*}
\delta\mathcal{J}_{d5}&=\langle\lambda^5_k,\dfrac{\textrm{ad}_{X_{uk}}\exp(\textrm{ad}_{X_{uk}})}{\exp(\textrm{ad}_{X_{uk}})-1}(\eta_{uk+1}-\textrm{Ad}_{f^{-1}_{uk}}\eta_{uk}-\chi_{uk})\rangle\\
&\textrm{Setting }\lambda^5_k\equiv\left(\dfrac{\textrm{ad}_{X_{uk}}\exp(\textrm{ad}_{X_{uk}})}{\exp(\textrm{ad}_{X_{uk}})-1}\right)^*\lambda^5_k,\\
&=\langle \lambda^5_k, \eta_{uk+1}-\textrm{Ad}_{f^{-1}_{uk}}\eta_{uk}-\chi_{uk}\rangle\\
&=\langle \lambda^5_k, \eta_{uk+1}\rangle+\langle-\textrm{Ad}^*_{f^{-1}_{uk}}\lambda^5_k,\eta_{uk}\rangle+\langle-\lambda^5_k,\chi_{uk}\rangle
\end{align*}
\normalsize
Using the fact that the end-points are fixed, (\ref{vj1}) gives us the required necessary conditions for all admissible variations and hence, completes the proof.\hfill$\blacksquare$\\\\

\end{document}